\documentclass[12pt,a4paper]{article}%

\usepackage{cite}
\usepackage{amsmath,amssymb,amsfonts}
\usepackage{calc}
\usepackage{hyperref}
\usepackage[pdftex]{color,graphicx}

\usepackage{pifont}% http://ctan.org/pkg/pifont
\newcommand{\cmark}{\ding{51}}%
\newcommand{\xmark}{${\bf \large  -}$}

\usepackage{multirow}

\usepackage{xcolor,colortbl}

\definecolor{Gray}{gray}{0.95}
\definecolor{RGray}{gray}{0.85}
\definecolor{CGray}{gray}{0.92}

\newcolumntype{a}{>{\columncolor{Gray}}c}
\newcolumntype{b}{>{\columncolor{white}}c}

\numberwithin{equation}{section} 
\newcommand{\eval}[2]{  $#1 \times 10^{-#2}$    }

\newcommand{\be}{\begin{equation}}
\newcommand{\ee}{\end{equation}}

\def\beqn{\begin{eqnarray}}
\def\ba{\begin{array}{c}}
\def\bat{\begin{array}{cc}}
\def\bat{\begin{array}{cc}}
\def\ea{\end{array}}
\def\bat{\begin{array}{cc}}
\def\batt{\begin{array}{ccc}}
\def\eeqn{\end{eqnarray}}
\def\cL{{\cal L}}

\definecolor{nicered}{rgb}{1.0,0.0,0.2}

\usepackage{color}

\begin{document}
\vspace*{-2.5cm}
\begin{flushright}
{\small
FTUV/14-0321\\
IFIC/14-20\\
LA-UR-14-21868\\
}
\end{flushright}
\renewcommand{\thefootnote}{\fnsymbol{footnote}}
\vspace{0.5cm}
\begin{center}

{\LARGE \bf   The model-discriminating power  of \\[5mm] lepton flavor violating  $\tau$ decays}  \\

\bigskip\color{black}\vspace{0.6cm}
{{\large \bf Alejandro Celis$^{a}$, Vincenzo Cirigliano$^{b}$ and Emilie Passemar$^{b,}$\footnote{Email: \url{alejandro.celis@ific.uv.es},~ \url{cirigliano@lanl.gov},~ \url{passemar@lanl.gov}}}
} \\[7mm]
{\it  (a)  IFIC, Universitat de Val\`encia -- CSIC, Apt. Correus 22085, E-46071 Val\`encia, Spain}\\[3mm]
{\it  (b)  Theoretical Division, Los Alamos National Laboratory, Los Alamos,  NM 87545, USA}\\[3mm]
\end{center}
\bigskip
\centerline{\large\bf Abstract}
\begin{quote}

Within an effective field theory framework, we discuss the possibility to discriminate among different  operators that contribute to lepton flavor violating (LFV)  $\tau$ decays.    Correlations among decay rates in different channels are shown to provide a basic handle to unravel the origin of LFV in these processes.  More information about the underlying dynamics responsible for LFV can be gathered from differential distributions in three-body decays like $\tau \rightarrow \mu \pi \pi$ or $\tau \rightarrow 3 \mu$:  these are considered in some detail.    We incorporate in our analysis recent developments in the determination of the hadronic form factors for $\tau \rightarrow \mu \pi \pi$.     Future prospects for the observation of LFV $\tau$ decays and its interpretation are also discussed. 

\end{quote}

\renewcommand{\thefootnote}{\arabic{footnote}}\setcounter{footnote}{0}

\pagestyle{empty}

\newpage
\section{Introduction}
 \setcounter{page}{1}
\pagestyle{plain}
The observation of charged lepton flavor violation (CLFV) would be a clear indication of physics beyond the Standard Model (SM), see Ref.~\cite{Raidal:2008jk} for a comprehensive review.    Moreover, the search for CLFV is complementary to  new physics (NP) searches at the energy frontier as well as with other areas of the intensity  frontier program (rare $B$ and $K$ decays, electric dipole moments, the muon anomalous magnetic moment, among others).  Many scenarios of physics beyond the SM predict rates for LFV  processes  
of charged leptons within the  reach  of present and future experiments.    Some examples are: the SM with additional right-handed heavy Majorana neutrinos or with left-handed and right-handed neutral singlets~\cite{Cvetic:2002jy}, supersymmetric  
models~\cite{Borzumati:1986qx,Barbieri:1995tw,Hisano:1995cp,Ellis:2002fe,Dedes:2002rh,Masiero:2002jn,Fukuyama:2003hn,Brignole:2004ah,Arganda:2008jj,Herrero:2009tm,Hirsch:2012ax,Altmannshofer:2013lfa}, 
left-right symmetric models~\cite{Barenboim:1996vu,Cirigliano:2004mv}, 
technicolor models with non-universal $Z^{\prime}$ exchange~\cite{Yue:2002ja}, multi-Higgs doublet models~\cite{Diaz:2000cm,Kanemura:2005hr,Paradisi:2005tk,Davidson:2010xv,Crivellin:2013wna,Celis:2013xja}, leptoquark models~\cite{Davidson:1993qk,Gabrielli:2000te,Arnold:2013cva}, models with heavy vector-like leptons~\cite{McKeen:2012av,Redi:2013pga,Ishiwata:2013gma,Falkowski:2013jya}, and, the Littlest Higgs model with $\mathrm{T}$-parity~\cite{delAguila:2008zu,Blanke:2009am}.    It is obvious then that if LFV transitions among charged leptons are observed at some point, it will be a challenging task to disentangle all the possible NP candidates.   In this work we discuss the issue of discriminating NP contributions in LFV $\tau$ decays. 

The set of LFV searches that can be  performed with the $\tau$-lepton is very different from those in the  $\mu - e$ sector.   Searches for LFV at low energy are being pursued  in $\mu$ to $e$ conversion in nuclei, radiative $\mu \rightarrow e \gamma$ and leptonic $\mu \rightarrow 3 e$ decays.   The relatively heavy mass of the $\tau$ lepton compared with that of the lightest hadrons, opens a rich variety of LFV semileptonic $\tau$ decay modes $\tau \rightarrow \ell\, (\pi, \eta^{(\prime)}, \pi \pi, \ldots)$.    Together with radiative $\tau \rightarrow \ell \gamma$ and leptonic $\tau \rightarrow \ell \ell^{\prime}  \bar \ell^{\prime \prime} $ decays, semileptonic decays offer an interesting window to probe the underlying LFV mechanism, being particularly sensitive to different kinds of NP or effective operators.

Current bounds on LFV $\tau$ decay rates have been set by the Belle and BaBar collaborations, improving considerably over previous limits.  The LHCb collaboration has presented recently a search for $\tau \rightarrow 3 \mu$ decays, obtaining an upper limit which is already comparable with that of $\mathrm{B}$-factories~\cite{Aaij:2013fia}.   In the near future, the Belle II experiment at the SuperKEKB collider will bring the search for NP effects associated with the $\tau$-lepton to a new level of sensitivity.  It is expected that a sensitivity gain of an order of magnitude can be achieved in many LFV $\tau$ decay modes with $50$~ab$^{-1}$ of collected data~\cite{Abe:2010sj}.  The possibility of a future Super Tau-Charm Factory could also bring important improvements on  the  sensitivity to LFV $\tau$ decays, reducing considerably the large background from $e^+ e^- \rightarrow \tau^+ \tau^- \gamma$ compared to $\mathrm{B}$-factories~\cite{Bobrov:2012kg}. 

If LFV transitions are observed at some point, a comparison among the measured rates as well as upper limits on other non-observed processes, will provide information to discriminate among possible NP models.  Detailed treatments of LFV $\mu - e$ transitions within an effective field theory (EFT) approach to  NP have been performed in Refs.~\cite{Raidal:1997hq,Kuno:1999jp,Kitano:2002mt,Cirigliano:2009bz}.  From these works it emerges that the nucleus-dependence of  $\mu$ to $e$ conversion rate is a powerful diagnosing tool, that 
extends the discriminating power of $\mu \to e \gamma$ and $\mu \to 3 e$~\cite{deGouvea:2000cf} to operators involving quarks.

 Phenomenological analyses of LFV $\tau$-decays within a generic EFT framework have also discussed the discrimination of NP scenarios by comparing the decay rates in different $\tau$-decay modes~\cite{Black:2002wh,Brignole:2004ah,Dassinger:2007ru,Matsuzaki:2007hh,Giffels:2008ar,Petrov:2013vka,Crivellin:2013hpa}.   More information can be gathered by studying in detail the differential distributions in three-body decays.  For the leptonic decays, $\tau \rightarrow \ell \ell^{\prime}  \bar \ell^{\prime \prime}$, a Dalitz plot analysis can be used to distinguish different types of effective operators~\cite{Dassinger:2007ru,Matsuzaki:2007hh}.     In semileptonic $\tau \rightarrow \ell \pi \pi$ decays on the other hand, the pion invariant mass spectrum offers an alternative tool to separate different NP contributions.    In most of the previous works treating these decays, the determination of the scalar and gluonic hadronic matrix elements has been based on leading-order Chiral Perturbation Theory (ChPT) predictions~\cite{Black:2002wh,Kanemura:2005hr,Arganda:2008jj,Petrov:2013vka}.    The problem in such a description lies in the fact that the invariant mass of the pion pair in these decays can be relatively large $\sqrt{s} < m_{\tau} - m_{\ell}$, while ChPT is only reliable at very low energies (well below the $\rho$ mass).  

 A proper treatment of the hadronic matrix elements for $\tau \rightarrow \ell \pi \pi$ decays was given for the first time in Refs.~\cite{Daub:2012mu,Celis:2013xja} within the framework of $\mathrm{R}$-parity violating supersymmetry and extended Higgs sectors, respectively.   The form factors derived in these works however can be used generically in other NP scenarios.       In this work, we adopt a general approach to describe LFV $\tau$ decays.  The information of possible new heavy degrees of freedom is encoded in an effective Lagrangian which describes the relevant physics at the low energy scale of the $\tau$-lepton mass.  This allows us to analyze the sensitivity to different types of NP  in which  particular operators are expected to provide the dominant effects without resorting to specific details of the UV dynamics.  Special attention is given to semileptonic $\tau \rightarrow \mu \pi \pi$ decays for which considerable improvement over previous works is made, thanks to a proper determination of the form factors in the resonance region, covering essentially all the accessible phase space.

 In Sec.~\ref{lageff} we present the effective Lagrangian used to analyze LFV $\tau$ decays in this work.   The decay rates for the processes considered can be found in Sec.~\ref{decayratessec}.    In Sec.~\ref{phenosec} we discuss how the observation of patterns among different $\tau$ decay rates as well as analyses of differential distributions for three-body decays can be used to probe the underlying source of LFV.   Future prospects for the observation of LFV $\tau$ decays are analyzed in Sec.~\ref{sec:future}.  The conclusions of our work are given in Sec.~\ref{sum}.     A brief discussion of the hadronic matrix elements relevant to describe LFV semileptonic $\tau$ decays is relegated to Appendix~\ref{hadronic}, we refer the reader to Refs.~\cite{Daub:2012mu,Celis:2013xja} for more details.

\section{Effective Lagrangian at low energy for LFV $\tau$ decays}   \label{lageff}
We assume there is an energy scale $\Lambda \gg m_{\tau}$  at which sizable LFV effects are generated.     Let us  consider three frameworks for physics beyond the SM which give rise to the effects we will be interested in this work:

\begin{description}

\item[(a)]  A general two-Higgs doublet model (2HDM) without large-mass decoupling.\footnote{The term large-mass decoupling has been coined for the first time in Ref.~\cite{Haber:2013mia}.  In the Higgs basis where only one doublet is responsible for the EWSB, $\langle H_1^0 \rangle = v/\sqrt{2}$ and $\langle H_2^0 \rangle = 0$ with $v \simeq 246$~GeV, the large-mass decoupling limit occurs when the coefficient of the quadratic term $\mu_2 H_2^{\dag} H_2$ in the scalar potential satisfies $\mu_2 \gg v^2$ and quartic scalar couplings remain perturbative.  }  The energy scale of the dynamics responsible for LFV corresponds to the EW scale in this case $\Lambda \sim v \simeq 246$~GeV.

\item[(b)] The SM as a low energy effective theory in which heavy particles belonging to the UV completion of the SM (SUSY partners for example) have been integrated out.   One assumes in this case that there is a mass gap between the EW scale and the scale $\Lambda \gg v$ where new heavy particles appear.

\item[(c)] The SM, including a light scalar boson $h$, as an effective theory at the weak scale of some unknown dynamics of electroweak symmetry breaking.  The EW symmetry is assumed to be non-linearly realized in Nature.  The EW symmetry breaking scale $\Lambda$ is taken to be at $4 \pi v$ or above.

\end{description}

In the scenario (a) of the general 2HDM, we are interested here in the case where the model is not in the large-mass decoupling limit~\cite{Haber:2013mia} and all the scalars lie at the EW scale.  The case of large-mass decoupling can be regarded as a specific realization of our second scenario (b).   The general 2HDM contains tree-level flavor-changing neutral currents (FCNCs) in the Higgs sector leading to Higgs mediated LFV transitions~\cite{Diaz:2000cm,Kanemura:2005hr,Paradisi:2005tk,Davidson:2010xv,Crivellin:2013wna,Celis:2013xja}.     At the energy scale relevant to describe $\tau$ decays $E \sim m_{\tau}$, one can integrate out the heavy fields arriving then to LFV effective operators suppressed by the scalar masses $\Lambda \sim M_{\varphi}$.  For example,  the following four-fermion operators are generated due to scalar exchange
\be
  \frac{1}{M_{\varphi}^2}  \left( \bar \ell \, (1 \pm \gamma_5 )   \, \tau \cdot    \bar q  \{ 1, \gamma_5\} q  \right),
\ee
where $q$ denotes a light quark.  

In scenario (b) one assumes that there is an energy gap between the EW scale and the scale of NP beyond the SM denoted by $\Lambda \gg v \sim 246$~GeV.     The SM including the Higgs doublet $H$, is considered as an effective low-energy theory valid at the EW scale and EWSB occurs due to the non-vanishing expectation value of the Higgs doublet $\langle H^0 \rangle = v/\sqrt{2}$.   After integrating out the heavy degrees of freedom at the scale $\Lambda \gg v \sim 246$~GeV one arrives to the following SM effective Lagrangian
\be
\mathcal{L}_{\mathrm{SM}} = \mathcal{L}_{\mathrm{SM}}^{(4)} + \frac{1}{\Lambda} \sum_k\, C_{k}^{(5)} Q_k^{(5)} + \frac{1}{\Lambda^2} \sum_k \,C_{k}^{(6)} Q_k^{(6)} + \mathcal{O}\left(\frac{1}{\Lambda^3}\right) \,.
\ee
Here $\mathcal{L}_{\mathrm{SM}}^{(4)}$ stands for the renormalizable SM Lagrangian, the higher dimensional effective operators $Q_{k}^{(n)}$ are built with the SM degrees of freedom and are invariant under the SM gauge group $\mathrm{SU(3)}_C \times \mathrm{SU(2)}_L \times \mathrm{U(1)}_Y$.  The Wilson coefficients $C_{k}^{(n)}$ are dimensionless constants which encode NP effects due to the UV dynamics. At dimension five there is only the well known Weinberg operator\cite{Weinberg:1979sa}.  The basis of dimension-six effective operators has been obtained in Refs.~\cite{Buchmuller:1985jz,Grzadkowski:2010es}.     
The basis of baryon-number conserving dimension-six operators in the SM effective Lagrangian consists of 59 independent operators (barring flavor structure and Hermitian conjugations).   Considering the flavor indices, the dimension-six Lagrangian grows considerably and contains 2499 hermitian operators and real parameters~\cite{Alonso:2013hga}.  Many of these operators are lepton flavor violating.

In the last framework  we have considered (c), the EW symmetry is assumed to be non-linearly realized and a Higgs-singlet field is introduced in the spectrum to account for the new boson with mass around $126$~GeV; see Refs.~\cite{Alonso:2012pz,Pich:2012dv,Contino:2013kra,Buchalla:2012qq,Buchalla:2013rka} for recent discussions along this direction.     The next-to-leading-order (NLO) effective Lagrangian of the SM with a dynamically broken EW symmetry takes the form~\cite{Buchalla:2012qq,Buchalla:2013rka}\footnote{Here we have neglected the custodial-symmetry breaking term  $\mathcal{L}_{\beta_1}$ included in Ref.~\cite{Buchalla:2013rka}. }
\be
\mathcal{L } = \mathcal{L}_{LO} + \sum_i c_i \frac{v^{6-d_i}}{\Lambda^2}  \, \mathcal{O}_i \,, 
\ee
where $d_i$  is  the operator dimension and $\mathcal{L}_{LO}$ represents the leading-order effective Lagrangian, which is in general non-renormalizable.  The EW symmetry breaking scale $\Lambda$ is taken to be at $4 \pi v$ or above.     Among the effective operators $\mathcal{O}_i$ of the NLO effective Lagrangian one encounters for example four-fermion operators that violate lepton flavor in general, see Refs.~\cite{Buchalla:2012qq,Buchalla:2013rka} for a complete list of such operators.

In this work we are interested in performing a general description of LFV $\tau$ decays within the EFT language.  All the above weak-scale scenarios ultimately match onto a low-energy effective theory. Here
we assume that there are no light particles in the spectrum (like axions or sterile neutrinos) beyond those already discovered. 
Therefore, the relevant degrees of freedom are  the leptons ($e, \mu, \tau$), the light quarks ($u,d,s$) together with the gluon and photon gauge fields.  One should then build the most general effective Lagrangian with these degrees of freedom, keeping the invariance under the Lorentz symmetry and the unbroken $ \mathrm{SU(3)}_C \times \mathrm{U(1)}_{\mathrm{em}} $ gauge symmetry.  We will restrict the discussion of LFV transitions to the $\tau - \mu$ sector, all the results found in this work can be extrapolated to the $\tau - e$ sector in a trivial manner.
All our statements below apply to the low-scale ($\mu \sim 2$~GeV) Wilson coefficients.  These are related to ultraviolet physics by a matching calculation at the new physics scale and 
the appropriate renormalization group evolution, including additional threshold effects associated with integrating out the $W$ and $Z$ bosons, the Higgs, and heavy quarks.

The general low-scale effective Lagrangian describing LFV $\tau-\mu$ transitions
can be organized according to the type of operators present:
\begin{align}  \label{taulag}
\cL_{eff} &=  \cL_{eff}^{(D)} + \cL_{eff}^{(\ell q)}  + \cL_{eff}^{(G)} +  \cL_{eff}^{(4 \ell)}  + \cdots \,,
\end{align}
where the dots stands for operators of higher dimension.    Here $\cL_{eff}^{(D)}$ contains the effective dipole operators of dimension five
\begin{align}  \label{taulag1}
\cL_{eff}^{(D)} &=  - \frac{m_{\tau}}{\Lambda^2}  \,   \Bigl\{ \,   \, \left( \mathrm{C_{DR}}  \, \bar \mu \, \sigma^{\rho \nu} \, P_L \, \tau 
+ \mathrm{C_{DL}} \, \bar \mu \, \sigma^{\rho \nu} \, P_R \, \tau \right) F_{\rho \nu}  + \mathrm{h.c.}  \Bigr\} \,,
\end{align}
while the dimension-six four-fermion operators involving two quark fields are grouped in $\cL_{eff}^{(\ell q)}$,
\begin{align}  \label{taulag2} 
\cL_{eff}^{(\ell q)} &=   - \frac{1}{\Lambda^2}  \,  \sum_{q=u,d,s}  \Bigl\{    \left( \mathrm{C^{q}_{VR}} \, \bar \mu \,  \gamma^{\rho} \,  P_R \, \tau  \; + \mathrm{C^{q}_{VL}}  \, \bar \mu  \, \gamma^{\rho} \, P_L \, \tau  \;\right)    \bar q \, \gamma_{\rho} \,  q   \nonumber \\ 
&+ \left( \mathrm{C^{q}_{AR}} \,   \bar \mu \, \gamma^{\rho} \,  P_R \, \tau  + \mathrm{C^{q}_{AL}} \,   \bar \mu \, \gamma^{\rho} \, P_L \, \tau \right)   \bar q \, \gamma_{\rho} \gamma_{5} \, q \nonumber \\ 
&+  m_{\tau} m_{q} G_F\,   \left(  \mathrm{C^{q}_{SR}} \, \bar \mu \, P_L \, \tau     + \mathrm{C^{q}_{SL}} \bar \mu  \, P_R \, \tau  \right)   \bar q \,q \nonumber \\  
&+  m_{\tau} m_{q} G_F\,  \left(   \mathrm{C^{q}_{PR}}  \, \bar \mu \, P_L \, \tau   + \mathrm{C^{q}_{PL}}  \,  \bar \mu \, P_R \, \tau  \right)   \bar q \, \gamma_{5}\, q  \nonumber  \\  
&+  m_{\tau} m_{q} G_F\,   \left( \mathrm{C^{q}_{TR}}  \,   \bar \mu \, \sigma^{\rho \nu}  P_L \, \tau  + \mathrm{C^{q}_{TL}}  \,  \bar \mu \, \sigma^{\rho \nu} \,  P_R\,  \tau     \right)  \bar q \, \sigma_{\rho \nu}\,  \, q  
+ \mathrm{h.c.} \Bigr\}  \,.
\end{align}
Effective gluonic operators of dimension-seven are contained in $\cL_{eff}^{(G)}$,
\begin{align}  \label{taulag3}
\cL_{eff}^{(G)} &=   - \frac{ m_{\tau} G_F }{\Lambda^2}  \frac{\beta_L}{4 \alpha_s}  \,   \Bigl\{   \, \left(  \mathrm{C_{GR}}  \,  \bar \mu \, P_L \, \tau + \mathrm{C_{GL}} \, \bar \mu \, P_R \,  \tau   \right)   \, G_{\rho \nu}^{a} G_{a}^{\rho \nu}  \nonumber \\ 
&+\,  \left( \mathrm{C_{ \widetilde{G}R}}  \,  \bar \mu \, P_L \, \tau + \mathrm{C_{ \widetilde{G}L}}  \, \bar \mu \, P_R \,  \tau   \right)  \,    G_{\mu \nu}^{a} \widetilde{G}_{a}^{\mu \nu}
+ \mathrm{h.c.} \Bigr\}  \,,
\end{align}
with $\beta_L/(4 \alpha_s) = -9 \alpha_s/(8 \pi)$.    Note that for the previous operators in Eqs.~\eqref{taulag1}, \eqref{taulag2} and \eqref{taulag3}, the chirality of the Wilson coefficient corresponds to the chirality of the final muon in a generic decay $\tau \rightarrow \mu X$.  The last part, $\cL_{eff}^{(4 \ell)}$, includes the effective four-lepton operators
\begin{align}  \label{taulag4}
\cL_{eff}^{(4 \ell)} &=     - \frac{1}{\Lambda^2}  \,  \Bigl\{    \mathrm{C_{SLL}}  \,  \left( \bar  \mu \, P_L \, \tau  \right) \left( \bar  \mu \, P_L \, \mu \right)  + \mathrm{C_{SRR}}  \,  \left(  \bar \mu \, P_R \,  \tau \right)  \left(  \bar  \mu \,  P_R \, \mu \right)  \nonumber  \\  
 &+   \mathrm{C_{VLL}}  \,  \left(   \bar \mu   \gamma^{\mu} \, P_L \,  \tau \right)   \left( \bar \mu  \, \gamma_{\mu}  \,  P_L \,  \mu  \right)     +  \mathrm{C_{VRR}} \,  \left(  \bar \mu \,  \gamma^{\mu}   \,  P_R  \, \tau\right)  \left(  \bar  \mu   \, \gamma_{\mu} \,  P_R \, \mu  \right) \nonumber  \\  
 &+ \mathrm{C_{VLR}}  \,  \left( \bar  \mu \,  \gamma^{\mu} \,  P_L  \, \tau  \right) \left(  \bar   \mu \,  \gamma_{\mu} \,  P_R \,  \mu \right)   +  \mathrm{C_{VRL}} \,  \left( \bar \mu \,  \gamma^{\mu} P_R \,  \tau \right) \left(  \bar  \mu    \,\gamma_{\mu} \, P_L \, \mu \right)  +   \mathrm{h.c.} \Bigr\}  \,.
\end{align}
For simplicity we will consider only the leptonic decay mode $\tau \rightarrow 3 \mu$ in this work, other leptonic decay channels as $\tau^- \rightarrow e^- \mu^+ \mu^-$ will involve similar operators to those in Eq.~(\ref{taulag4}) but with independent Wilson coefficients in general.    We use $P_{R,L} = (1\pm\gamma_5)/2$, $\sigma^{\rho \nu} = \frac{i}{2} \left[ \gamma^{\rho}, \gamma^{\nu} \right]$ and  $G_F = (\sqrt{2} v^2)^{-1}$ for the Fermi constant.   The photon and gluon field strength tensors are denoted by $F_{\rho \nu }$ and $G^{a}_{\rho \nu}$ respectively.     The dual tensor of the gluon field strength is defined by  $\widetilde G^{a}_{\rho \nu} = \frac{1}{2}\, \epsilon_{\rho \nu \alpha \beta}  \,G^{a,\, \alpha \beta}$.    In the following we assume that $\mathrm{C_{TL}^{q}} = \mathrm{C_{TR}^{q}}  = 0$ and neglect higher-dimensional operators in the effective Lagrangian.   Since we are not  interested in CP-violating effects we will take all the Wilson coefficients to be real.

%%%%%%%%%%%%%%%%%%%%%%%%%%%%%%%%%%%%%%%%%%%%%%%%%%%%%%%%%%%%%%%%%%%%%%%%%%
\begin{table}[t]
\setlength{\extrarowheight}{3pt}
\begin{center}
\caption{\label{tab::dataP} \it \small Experimental upper bounds for LFV $\tau$ decays.  }
\vspace{0.4cm}
\doublerulesep 0.8pt \tabcolsep 0.4in
\small{
\begin{tabular}{|clc|c|}\hline\hline
\rowcolor{RGray}
$\tau^-$ decay mode                                   & Upper bound on $\mathrm{BR}$   ($90~\%$~CL)                                &     Comment \\ \hline
$ \mu \,\gamma $                                            & $ 4.4 \times10^{-8} $           &  \cite{Beringer:1900zz,Aubert:2009ag}   \\
\rowcolor{Gray}
$ \mu^- \, \mu^+ \mu^- $                                &$ 2.1 \times10^{-8} $            &  \cite{Beringer:1900zz,Hayasaka:2010np}  \\
$ \mu \,\pi^0 $                                &$ 1.1 \times10^{-7} $            &  \cite{Beringer:1900zz,Aubert:2006cz}  \\
\rowcolor{Gray}
        $\mu \, \eta$      &   $6.5 \times 10^{-8}$              & \cite{Beringer:1900zz,Miyazaki:2007jp}  \\ 
          $\mu \, \eta^{\prime}$      &    \eval{1.3}{7}         & \cite{Beringer:1900zz,Miyazaki:2007jp}  \\ 
          \rowcolor{Gray}
          $\mu\, \pi^+ \pi^-  $                                      &  \eval{2.1}{8}                  &  \cite{Miyazaki:2012mx} \\ 
                 $\mu \, \rho $                                      &  $1.2 \times 10^{-8}    $                &  \cite{Beringer:1900zz,Miyazaki:2011xe} \\
                 \rowcolor{Gray}
          $\mu \, f_0$   & \eval{3.4}{8} & \cite{Beringer:1900zz,Miyazaki:2008mw}  \\
\hline\hline
\end{tabular}}
\end{center}
\end{table}
%%%%%%%%%%%%%%%%%%%%%%%%%%%%%%%%%%%%%%%%%%%%%%%%%%%%%%%%%%%%%%%%%%%%%%%%%%

%%%%%%%%%%%%%%%%%%%%%%%%%%%%%%%%%%%%%%%%%%%%%%%%%%%%%%%%%%%%%%%%%%%%%%%%%%
%%%%%%%%%%%%%%%%%%%%%%%%%%%%%%%%%%%%%%%%%%%%%%%%%%%%%%%%%%%%%%%%%%%%%%%%%%
\begin{table}[t]
\setlength{\extrarowheight}{3pt}
\begin{center}
\caption{\label{tab::sensitivityP} \it \small Sensitivity of LFV $\tau$ decays to the different effective operators at tree-level.   The symbol  \cmark~(\xmark) denotes that the operator does (not) contribute at tree-level to a given process.    For operators involving 
quark bilinears, the relevant isospin structure ($I=0,1$) probed by a given decay  is also specified.  
}
\vspace{0.4cm}
\doublerulesep 0.8pt \tabcolsep 0.1in
\small{
\begin{tabular}{|ccccccc|}\hline\hline
\rowcolor{RGray}
                                                                &              $\tau \rightarrow 3 \mu$                       &    $\tau \rightarrow \mu \gamma$         &               $\tau \rightarrow \mu \pi^+ \pi^-$   &  	$\tau \rightarrow \mu K \bar{K} $    &            $\tau \rightarrow \mu \pi$          &             $\tau \rightarrow \mu \eta^{(\prime)}$  
                                                                \\ \hline 
 %%%%%% %%%%%% %%%%%% %%%%%% %%%%%% %%%%%% %%%%%% %%%%%% %%%%%% %%%%%% %%%%%% %%%%%% %%%%%% %%%%%% %%%%%% %%%%%% %%%%%% %%%%%% %%%%%% %%%%%% %%%%%% %%%%%% %%%%%%
$  \mathrm{C_{SLL,RR}} $                  & 				  \cmark             	              &    		 \xmark			     & 		  \xmark					 &					\xmark			  &		 		    \xmark 			    & 			   \xmark   					   				 
 \\  \rowcolor{Gray}
$  \mathrm{C_{VLL,RR}} $                  &  				 \cmark 		              & 			    \xmark 			     &     			     \xmark  		    &   				\xmark			 &				    \xmark    		     &				    \xmark 							 
  \\
$  \mathrm{C_{VLR,RL}} $ 		& 				  \cmark     	              &   		  \xmark 			     & 				     \xmark			    & 					\xmark			  & 				    \xmark  			    & 				     \xmark   						
 \\  \rowcolor{Gray}  
 %%%%%% %%%%%% %%%%%% %%%%%% %%%%%% %%%%%% %%%%%% %%%%%% %%%%%% %%%%%% %%%%%% %%%%%% %%%%%% %%%%%% %%%%%% %%%%%% %%%%%% %%%%%% %%%%%% %%%%%% %%%%%% %%%%%% %%%%%%
$\mathrm{C_{DL,R}}$                          &                                 \cmark     		                 &		  \cmark				 &  				  \cmark		              &				\cmark		   	    &			\xmark  		      	        &			    \xmark  					 
\\ 
$ \mathrm{C^{q}_{VL,R}}$    	         &   				\xmark      			       &  		   \xmark			          &		    	    \cmark  \, (I=1) 	    &	    			\cmark (I=0,1)         &		   \xmark 				     & 			 \xmark 	 				     
 \\   \rowcolor{Gray}  
$ \mathrm{C^{q}_{SL,R}}$                   & 				  \xmark            		      &  		   \xmark 			        & 			     \cmark  \, (I=0)   	    &				\cmark (I=0,1)	     &		   	\xmark   				     & 			   \xmark 					     
 \\ 
$\mathrm{C_{GL,R}}$                          & 				  \xmark 		              & 			    \xmark 			    &  			  \cmark				            &	 			\cmark			    &     		 \xmark    				    &  			  \xmark 					   	
  \\    \rowcolor{Gray}  
$\mathrm{C^{q}_{AL,R}}$      		&				 \xmark     			       &		   \xmark        			& 				     \xmark     	     	  &				\xmark			 	&		 \cmark   \, (I=1)     	     &			  \cmark  \, (I=0) 	              
   \\
$  \mathrm{C^{q}_{PL,R}} $               & 				\xmark 		              &			    \xmark     		         &				  \xmark		   	&				\xmark			    & 	 	        \cmark   \, (I=1) 		     &  			\cmark    \, (I=0)            	     
  \\   \rowcolor{Gray}  
$\mathrm{C_{ \widetilde{G}L,R}}$   & 				  \xmark			     &  		   \xmark 			    &  				  \xmark 		            &					\xmark			   &	   			  \xmark    			    &  			  \cmark   					   
 \\
% \rowcolor{Gray} 
\hline\hline	
\end{tabular}}
\end{center}
\end{table}
%%%%%%%%%%%%%%%%%%%%%%%%%%%%%%%%%%%%%%%%%%%%%%%%%%%%%%%%%%%%%%%%%%%%%%%%%%

%%%%%%%%%%%%%%%%%%%%%%%%%%%%%%%%%%%%%%%%%%%%%%%%%%%%%%%%%%%%%%%%%%%%%%%%%%

%
%
%
%
\section{Lepton flavor violating $\tau$ decays}  \label{decayratessec}
If LFV $\tau$ decays are observed at some point in the future, one would like to gain as much information as possible about the underlaying dynamics responsible for LFV.   This can be done for example by looking for correlations among different decay modes or by analyzing differential decay distributions in three-body $\tau$ decays, this will be discussed in Sec.~\ref{phenosec}.   We consider in this work radiative and leptonic LFV $\tau$ decays as well as semileptonic decay modes, for which the current experimental limits  are summarized in Table~\ref{tab::dataP}.
These decays are sensitive to specific combinations of effective operators as shown in Table~\ref{tab::sensitivityP}.    
If a  given  type of operator dominates one expects to observe a particular pattern for the branching ratios (BR) of the different decay channels.   
In this section we provide expressions for the LFV $\tau$ decay rates considered.

\subsection{Radiative and leptonic decays}
\label{sect:rad}

The radiative decay $\tau \rightarrow \mu \gamma$ receives contributions at tree-level only from the effective dipole operators in Eq.~\eqref{taulag}, the decay rate is given by
\be
\Gamma(\tau \rightarrow \mu \gamma )=   \frac{   m_{\tau}^5 }{   4 \pi \Lambda^4 }  \left(  |\mathrm{C}_{\mathrm{DL}}|^2 + |\mathrm{C}_{\mathrm{DR}}|^2  \right) \,,
\ee
where we have taken $m_{\mu} = 0$.    The LFV leptonic $\tau$ decay $\tau^{-}(p) \rightarrow \mu^-(p_1)  \mu^+(p_2) \mu^-(p_3)$ is sensitive to the effective dipole operators (connecting the photon to a $\mu^+\mu^-$ pair) and the four-lepton operators in Eq.~\eqref{taulag4}.  The doubly differential decay width can be written as
\begin{align}   \label{3mu}
\dfrac{d^2 \Gamma(\tau \rightarrow 3 \mu)}{  dm_{13}^2 dm_{23}^2 } \; =&\;   \frac{1}{1024 \pi^3 \Lambda^4  m_{\tau}^3 }  \Biggl\{  
 \frac{  64 \pi \alpha_{ \mathrm{em}}  m_{\tau}^2    }{     m_{23}^2   (   m_{13}^2 + m_{23}^2 - m_{\tau}^2   )  }  \Bigl[    -2 m_{\tau}^2   \left(    2 m_{13}^4    + 4 m_{13}^2 m_{23}^2 + m_{23}^4  \right) \nonumber \\ &+ 2 m_{13}^2 \left(   m_{13}^4  + 3 m_{13}^2  m_{23}^2 + 3 m_{23}^4 \right) +  m_{\tau}^4 \left(   3 m_{13}^2 + 2 m_{23}^2 \right)  - m_{\tau}^6  \Bigr] \,  |  \mathrm{C}_{  \mathrm{DL} }  |^2    \nonumber \\
&+ 4 \left[    m_{13}^2 (   m_{\tau}^2 - 2 m_{23}^2  )    +   2 m_{23}^2  (  m_{\tau}^2 -   m_{23}^2  )    -   m_{13}^4    \right]   \,  |\mathrm{C}_{\mathrm{VLR}}|^2   \nonumber  \\
& +   m_{13}^2   (   m_{\tau}^2 - m_{13}^2 ) \,  \left(  |\mathrm{C}_{\mathrm{SLL}}|^2  + 16 |\mathrm{C}_{\mathrm{VLL}}|^2 \right)    \nonumber \\
&+  32 (  \pi \alpha_{\mathrm{em}} )^{1/2} m_{\tau}^2  \,   \left[   4  m_{13}^2 \mathrm{C_{VLL}}  +(  m_{\tau}^2 - m_{13}^2  )  \mathrm{C_{VLR}}      \right] \mathrm{C_{DL}}  \nonumber \\
&+ (\mathrm{L} \leftrightarrow \mathrm{R})
  \Biggr\} \,.
\end{align}
% 
%Note that contributions mixing operators with different chiralities are absent in Eq.~\eqref{3mu} because we have taken $m_{\mu} = 0$. The invariant masses $m_{ij}^2 =   (  p_i + p_j)^2 $ are kinematically limited by:
Note that contributions arising from the interference of operators with different muon chirality  are absent in Eq.~\eqref{3mu} because we have taken $m_{\mu} = 0$. The invariant masses $m_{ij}^2 =   (  p_i + p_j)^2 $ are kinematically limited by:
 %
 %\begin{align}
 \begin{eqnarray}
4 m_{\mu}^2  &\leq & m^2_{13}  \leq    ( m_{\tau}  - m_{\mu} )^2  \\ 
\left(m^2_{23} \right)_{\mathrm{min}, \mathrm{max} } & =& \left(    E_2   + E_3 \right)^2 - \left(    \sqrt{   E_2^2  - m_{\mu}^2    }  \pm \sqrt{ E_3^2 -m_{\mu}^2      }  \right)^2   \,,
 \end{eqnarray}
% \end{align}
 %
 where 
\be
 E_2 =  \frac{  m_{\tau}^2  - m_{13}^2 - m_{\mu}^2 }{ 2  m_{13} } \,, \qquad \qquad E_3 =   \frac{m_{13}}{ 2 } \,
\ee
are the energies of $\mu^+(p_2)$ and $\mu^-(p_3)$ in the $m_{13}$ rest frame.   

\subsection{Semileptonic decays} 

Semileptonic $\tau$ decays are particularly useful to disentangle different effective operators. Quark bilinears have different $\mathrm{J^{PC}}$ quantum numbers, this implies that for some operators only a given set of hadronic final states is possible.       Semileptonic decays $\tau \rightarrow \mu P$ with a pseudoscalar meson in the final state probe pseudoscalar and axial four-fermion effective operators as well as LFV effective couplings with the Parity-odd gluonic operator $G_{\mu \nu}^{a} \widetilde{G}_{a}^{\mu \nu}$. For $\tau \rightarrow \mu \pi^0$, the decay width is given in the limit $m_{\mu} = 0$ by 
\begin{align} \label{eq:pi0}
\Gamma(\tau \rightarrow \mu \pi^0) \;=&\; \frac{   ( m_{\tau}^2 - m_{\pi}^2 )^2  }{  32 \pi\, m_{\tau} \Lambda^4 }  \Biggl\{      \left(  \mathrm{A}^{\pi}_{\mathrm{L}}   +  G_F \,  \mathrm{P}^{\pi}_{\mathrm{L}}      \right)^2     +   (\mathrm{L} \leftrightarrow \mathrm{R} )\Biggr\} \,,
\end{align}
with
\begin{align}
 \mathrm{A}^{\pi}_{\mathrm{L}} \;=\; (  \mathrm{C_{AL}^{u}}  - \mathrm{C_{AL}^{d}}   )  \frac{ f_{\pi}    }{ \sqrt{2} }   \,, \qquad 
 \mathrm{P}^{\pi}_{\mathrm{L}} \;=\; (  \mathrm{C_{PL}^{u}} - \mathrm{C_{PL}^{d}}   ) \frac{   m_{\pi}^2   }{   2 \sqrt{2} }  f_{\pi} \,.
\end{align}
The parameter $f_{\pi}$ corresponds to the pion decay constant and its numerical value is given in Appendix~\ref{hadronic}.  Similarly, the decay rate for $\tau \rightarrow \mu \eta$ can be written in the limit $m_{\mu} = 0$ as
\begin{align}   \label{mueta}
 \Gamma(\tau \rightarrow  \mu \eta)\; =\;&     \frac{(  m_{\tau}^2  - m_{\eta}^2 )^2}{ 32 \pi  m_{\tau} \Lambda^4   }   \Biggl\{  \left( \frac{ 9  \,  G_F a_{\eta} }{ 2} \right)^2   |\mathrm{C_{\widetilde{G}L}}|^2  +  \left(  \mathrm{A}^{\eta}_{\mathrm{L}}  +   G_F \,\mathrm{P}^{\eta}_{\mathrm{L}}  \right)^2     + (\mathrm{L}  \leftrightarrow \mathrm{R} )    \Biggr\}    \;,
\end{align}
where we have defined
\begin{align}
 \mathrm{A}^{\eta}_{\mathrm{L}}  \;=\;    (  \mathrm{C_{AL}^{u}} +\mathrm{C_{AL}^{d}}   )   \frac{f_{\eta}^q}{\sqrt{2}} + \mathrm{C_{AL}^{s}}    f_{\eta}^s \,, \qquad
 \mathrm{P}^{\eta}_{\mathrm{L}} \;=\; (  \mathrm{C_{PL}^{u}} + \mathrm{C_{PL}^{d}}   ) \frac{   h_{\eta}^{q} }{ 2 \sqrt{2}Ê}  + \mathrm{C_{PL}^{s}} \frac{h_{\eta}^{s} }{2} \,.  
\end{align}
The constants $\{ a_{\eta}, f_{\eta}^{q,s}, h_{\eta}^{q,s}\}$ parametrize the relevant hadronic matrix elements needed, see Appendix~\ref{hadronic} for their exact definition and their numerical values.   The relevant expression for $\Gamma(\tau \rightarrow \mu \eta^{\prime})$ can be obtained from Eq.~\eqref{mueta} via the  replacement $\eta \rightarrow \eta^{\prime}$.

Finally, the differential decay width for the semileptonic $\tau$ decay into a pair of charged pions $\tau \rightarrow \mu \pi^+ \pi^-$ can be written as
\begin{align}  \label{eqforpipi}
\frac{d \Gamma(\tau \rightarrow \mu \pi^+ \pi^-)}{ds} =&   \frac{   (  s - 4 m_{\pi}^2  )^{1/2}   (  m_{\tau}^2 - s )^2   }{ 1536 \pi^3 \, \Lambda^4 \, m_{\tau} \, s^{5/2} }  \nonumber \\
& \times \Biggl\{  3 s^2   G_F^2  | \mathrm{Q_L(s)} |^2 - 4 (4 m_{\pi}^2 - s ) | F_{V}(s) |^2  \Biggl[     4 \pi \alpha_{\mathrm{em}} (  2 m_{\tau}^2 + s )  |  \mathrm{C_{DL}} |^2  \nonumber \\
&+ s (   \mathrm{C_{VL}^d} - \mathrm{C_{VL}^u}   )   \Bigl(   12 \sqrt{  \pi \alpha_{\mathrm{em}}}   \, \mathrm{C_{DL}}  + \frac{(  m_{\tau}^2 + 2 s )}{m_{\tau}^2} (  \mathrm{ C_{VL}^{d}} - \mathrm{ C_{VL}^{u}} )     \Bigr) \Biggr]  \nonumber \\
&+ (\mathrm{L}  \rightarrow \mathrm{R})   \Biggr\}      \,.
\end{align}
Here we have taken $m_{\mu} = 0$ and
\begin{align}
\mathrm{Q_L}(s) \;=\;&  \Bigl(    \theta_{\pi}(s)    - \Gamma_{\pi}(s)  - \Delta_{\pi}(s)     \Bigr)     \mathrm{C_{GL}}   + \Delta_{\pi}(s) \,\mathrm{C_{SL}^{\,s}}  + \Gamma_{\pi}(s) \, \left( \mathrm{C_{SL}^{u}} + \mathrm{C_{SL}^{d}} \right) \,.
\end{align}
The invariant mass of the pion pair  $s = (p_{\pi^+} + p_{\pi^-} )^2$ is kinematically limited to $ 4 m_{\pi}^2  \leq  s  \leq (  m_{\tau} - m_{\mu} )^2 $.    The hadronic form factors $\{\Gamma_{\pi}(s), \Delta_{\pi}(s), \theta_{\pi}(s)\}$ and $F_{V}(s)$ are defined in Appendix~\ref{hadronic}.  The determination of these form factors was carried out recently in Refs.~\cite{Daub:2012mu,Celis:2013xja}.

There are also experimental bounds for semileptonic $\tau$ decays into a lepton and a short-lived resonance, as $\rho(770)$  $(\mathrm{J^{PC}} =1^{--})$ or $ f_0(980)$   $(\mathrm{J^{PC}} = 0^{++})$.   Bounds on the BR in this case are determined experimentally by applying a cut on the $\pi^+\pi^-$ invariant mass.  For $\rho(770)$ the cut is $ 587~\text{MeV} < \sqrt{s} < 962~\text{MeV}$~\cite{Miyazaki:2011xe}, while,  $  906~\text{MeV} < \sqrt{s} < 1065~\text{MeV}$ for  $f_0(980)$~\cite{Miyazaki:2008mw}.    In the following we will drop the mass label for these resonances and refer to them simply as $\rho$ and $f_0$.       Measurements for $\tau \rightarrow \mu \rho$ and $\tau \rightarrow \mu f_0$ decays probe different regions (though overlapping) of the pion invariant mass spectrum in $\tau \rightarrow \mu \pi^+ \pi^-$ decays.  A proper determination of the hadronic form factors in all the kinematical range is needed to extract meaningful information out of the experimental limits on $\tau \rightarrow \mu \pi^+ \pi^-, \mu \rho, \mu f_0$, see discussions in Refs.~\cite{Daub:2012mu,Celis:2013xja}.

In Table~\ref{tab::sensitivityP} we have included  for completeness the $\tau \rightarrow \mu K \bar K$ modes.  They are in principle quite useful because they are sensitive to all isospin structures 
for  both the scalar and vector operators.  This is not the case for $\pi \pi$ final states due to Bose statistics. 
Current knowledge of the relevant $K \bar{K}$ form factors, however, is not as firm as for the $\pi \pi$ modes.
The vector form factors can be obtained from Ref.~\cite{Arganda:2008jj}  and references therein. 
The scalar-isoscalar form factors  are in principle available from the analysis of Refs.~\cite{Daub:2012mu,Celis:2013xja}. 
Finally,  we are  not aware  of any determination of  the scalar-isovector form factor, 
although it  could be obtained in principle by a couple-channel dispersive analysis including the 
$K \bar{K}$ and $\pi \eta$ channels. 
In summary, we  drop the $K \bar{K}$ modes from our analysis 
due  to unknown or uncertain form factors,   smaller phase space, 
and worse experimental sensitivities compared to the $\pi \pi$ modes.

\section{Disentangling effective operators in LFV $\tau$ decays}  \label{phenosec}

We have two main handles to unravel the origin of LFV in $\tau$ decays. 
The first  is to look for correlations among the different LFV $\tau$ decay rates.  
For example,
if the dipole operator dominates over the remaining effective operators, we would expect to observe $\tau \rightarrow \mu \gamma$ before any other LFV $\tau$ decay.   Furthermore, the BR of those processes which also receive contributions from the dipole operator would be expected to be fixed relative to $\mathrm{BR}(\tau \rightarrow \mu \gamma)$, of course with some possible contamination due to contributions from other sub-leading operators.     Similar arguments can be formulated in case other type of operator(s) dominate.         
The second handle is provided by 
differential distributions in many-body decays, such as $\tau \to \mu \pi^+ \pi^-$ and $\tau \to 3 \mu$.
In this Section we discuss these  two handles in turn, after introducing a set of benchmark models.

\subsection{Benchmarks for the single operator dominance hypothesis}  \label{bench}

We will consider in the following a set of benchmark scenarios by assuming that only one type of operator is dominant.   For simplicity, we restrict the analysis to the case in which the outgoing muon has a definite chirality.    We will define benchmark scenarios relevant for the study of semileptonic LFV $\tau$ decays, leptonic decays like $\tau \rightarrow 3 \mu$ involve in general different effective operators (those in $\cL_{eff}^{(4 \ell)}$) and are discussed in   detail in Sec.~\ref{sst3mu}.

\begin{itemize}
\item {\bf{Dipole model}}

In the Dipole model one assumes that, among all the different effective operators, the dipole operator dominates.   Explicitly, we set in this scenario
\be 
\mathrm{C_D} \equiv  \mathrm{C_{DL}}   \neq 0 \,, \qquad \mathrm{C_{else}} = 0     \,.
\ee

\item 

{\bf{Scalar model}}

In this case we assume that the four-fermion scalar operator dominates and we take a Yukawa-like flavor  structure (recall that in the scalar operators we have pulled out an explicit factor of $m_q$):
\be
\mathrm{ C_S \equiv  C_{SL}^{u}  =  C_{SL}^{d}  = C_{SL}^{s}  }   \neq 0 \,, \qquad \mathrm{C_{else}} = 0     \,.
\ee

\item 

{\bf{Vector model  }}

This model is defined by:
\be 
\mathrm{C_{V^{(\gamma)}}}  \equiv  \mathrm{C_{VL}^{u}}  = -2   \mathrm{C_{VL}^{d}}   \neq 0 \,, \qquad \mathrm{C_{else}} = 0\,, 
\ee
with couplings proportional to the quark electric charges.

\item 

{\bf{$Z$-penguin model   }}

In this model we assume dominance of an effective Z-penguin LFV vertex. 
In this case, the Standard Model $Z$-fermion couplings fix the 
relative size of the Vector and Axial couplings as follows:
\begin{align}
\mathrm{C_{Z }}  \equiv  \mathrm{C_{VL}^{u} }            \,, \qquad \mathrm{C_{VL}^{d}}  = (v_d/v_u)     \mathrm{C_{VL}^{u}}     \,,
\end{align}
while the axial ones can be written as $\mathrm{C_{AL}^{q}}  =  - (a_q/v_u)   \mathrm{C_{VL}^{u}}$ with
\begin{align}
 v_u &= (1 - \frac{8}{3}  \sin^2 \theta_W )/2 \,,   \qquad  a_u = 1/2 \,,  \nonumber \\
 v_d &= ( - 1 +  \frac{4}{3}  \sin^2 \theta_W )/2 \,,   \, \qquad a_d =   -1/2  \, , 
\end{align}
where  $\sin^2 \theta_W \simeq 0.223$ is the weak mixing angle.

\item {\bf{Gluonic model (Parity-even)}}

In this model we consider only the Parity-even gluonic operator:
\be 
\mathrm{C_G}  \equiv  \mathrm{C_{GL}}  \neq 0 \,, \qquad \mathrm{C_{else}} = 0\,.
\ee

\item {\bf{Gluonic model (Parity-odd)}}

In this case only the Parity-odd gluonic operator is considered: 
\be 
\mathrm{C_{\widetilde G}}   \equiv \mathrm{C_{\widetilde GL}}   \neq 0 \,, \qquad \mathrm{C_{else}} = 0\,.
\ee

\item 

{\bf{Pseudoscalar model  1 }}

Four-fermion pseudoscalar operators are assumed to dominate with a Yukawa-like flavor structure, 
\be
\mathrm{ C_{P^{(1)} } \equiv  C_{PL}^{u}  =  C_{PL}^{d}  = C_{PL}^{s}  }   \neq 0 \,, \qquad \mathrm{C_{else}} = 0     \,.
\ee

\item 

{\bf{Pseudoscalar model  2 }}

In this case pseudoscalar operators are assumed to have a particular flavor structure: 
\be
\mathrm{ C_{P^{(2)}} \equiv  C_{PL}^{u}  =  -  C_{PL}^{d}  = - C_{PL}^{s}  }   \neq 0 \,, \qquad \mathrm{C_{else}} = 0     \,.
\ee

\end{itemize}

\subsection{Correlations between different $\tau$ decay modes}

To analyze correlations between different LFV $\tau$ decay modes in the single operator dominance hypothesis it will be useful to define the following ratio

\be
R_{F,M}  \equiv    \frac{\Gamma(\tau \rightarrow  F  )}{ \Gamma(\tau \rightarrow  F_M ) }  ~, 
\ee
where $F$ is a generic final state and $F_M$ represents the dominant LFV decay mode $\tau \rightarrow F_M$ in the model labeled by $M \in \{D, S, V^{(\gamma)}, Z, G, \tilde G,   P^{(1)}, P^{(2)}   \}$. 
For example in the Dipole scenario, $\mathrm{C_D} \neq 0$, the radiative decay mode dominates so that $F_{\mathrm{D}} = \mu \gamma$.    Within the single operator dominance hypothesis, all the dependence on the high energy scale $\Lambda$ and the Wilson coefficients cancels when taking the ratio.    
The patterns  of  
 $R_{F,M}$ in the different benchmark models are given in Tables~\ref{tab::patternsII} and \ref{tab::patternsI}.  We also provide limits on the BRs of the different decay modes in each scenario, extracted from the non-observation of LFV $\tau$ decays, using the current experimental upper bounds from Table~\ref{tab::dataP}.

In the Dipole model the dominant decay mode is $\tau \rightarrow \mu \gamma$, one obtains in this case
\be
\mathrm{BR}(\tau \rightarrow \mu \gamma) \simeq 6.2 \times 10^{11} \left(  \frac{  \mathrm{C_D} }{\Lambda^2} \right)^2 \, [\mathrm{GeV}^4] \,.
\ee
The strongest limit on the combination $\mathrm{C_D}/\Lambda^2$ is extracted from the experimental upper bound on $\mathrm{BR}(\tau \rightarrow \mu \gamma)$, giving
\be   \label{CDlimit}
\frac{|\mathrm{C_D}|}{\Lambda^2} < 2.7 \times10^{-10}~\text{GeV}^{-2}\,.
\ee
In the Scalar model on the other hand, the only decay channel is $\tau \rightarrow \mu \pi^+ \pi^-$ for which
\be
\mathrm{BR}(\tau \rightarrow \mu \pi^+ \pi^-) \simeq 1.9 \times 10^{-3} \left(  \frac{  \mathrm{C_S} }{\Lambda^2} \right)^2 \, [\mathrm{GeV}^4] \,.
\ee
In the Vector model, we have
\be
\mathrm{BR}(\tau \rightarrow \mu \pi^+ \pi^-) \simeq 4.3 \times 10^{9} \left(  \frac{  \mathrm{C_{V^{(\gamma)}} } }{\Lambda^2} \right)^2 [\mathrm{GeV}^4]  \,.
\ee
In the $Z$-penguin model the dominant decay mode is $\tau \rightarrow \mu \pi^+ \pi^-$: 
\be
\mathrm{BR}(\tau \rightarrow \mu \pi^+ \pi^-) \simeq 1.4 \times 10^{10} \left(  \frac{  \mathrm{C_{Z}}    }{\Lambda^2} \right)^2 [\mathrm{GeV}^4]  \,.
\ee
We have separated the $Z$-penguin model  in Tables~\ref{tab::patternsII} and \ref{tab::patternsI} for simplicity but it is important to note that in this case  the semileptonic modes $\tau \rightarrow \mu \pi^+ \pi^- $ and $\tau \rightarrow \mu \pi^0$ are related, the ratio $\Gamma(  \tau \rightarrow \mu \pi^+ \pi^- )/\Gamma(\tau \rightarrow \mu \pi^0) \simeq 2.8$ is fixed and does not depend on $\mathrm{C_{Z}}/\Lambda^2$.    Note that for the Vector and $Z$-penguin models the strongest bound on the relevant Wilson coefficient is extracted from $\tau \rightarrow \mu \rho$.   

In the Gluonic model (Parity-even) one obtains
\be
\mathrm{BR}(\tau \rightarrow \mu \pi^+ \pi^-) \simeq 0.02 \left(  \frac{  \mathrm{C_G} }{\Lambda^2} \right)^2 \, [\mathrm{GeV}^4] \,.
\ee
Only the semileptonic decays $\tau \rightarrow \mu P$ probe the Parity-odd Gluonic model, the dominant channel in this case is $\tau \rightarrow \mu \eta^{\prime}$,
\be
\mathrm{BR}(\tau \rightarrow \mu \eta^{\prime} ) \simeq 0.1 \left(  \frac{  \mathrm{ C_{  \widetilde{G} }} }{\Lambda^2} \right)^2 [\mathrm{GeV}^4]  \,.
\ee
For the Pseudoscalar models on the other hand
\be
\mathrm{BR}(\tau \rightarrow \mu \eta^{\prime} ) \simeq 2 \times 10^{-3} \left(  \frac{  \mathrm{ C_{ P^{(1)} }} }{\Lambda^2} \right)^2 [\mathrm{GeV}^4]  \,, \qquad \mathrm{BR}(\tau \rightarrow \mu \eta ) \simeq 2 \times 10^{-3} \left(  \frac{  \mathrm{ C_{ P^{(2)} }} }{\Lambda^2} \right)^2 [\mathrm{GeV}^4]   \,.
\ee
In the Pseudoscalar model 1, the strongest bound on the Wilson coefficient is obtained from the $\tau \rightarrow \mu \eta$ mode even though $\Gamma(\tau \rightarrow \mu \eta^{\prime}) > \Gamma(\tau \rightarrow \mu \eta)$.

%%%%%%%%%%%%%%%%%%%%%%%%%%%%%%%%%%%%%%%%%%%%%%%%%%%%%%%%%%%%%%%%%%%%%%%%%%
\begin{table}[t]
\setlength{\extrarowheight}{3pt}
\begin{center}
\caption{\label{tab::patternsII} \it \small  Expected pattern for the branching ratio of various LFV $\tau$ decays within the single operator dominance hypothesis.    }
\vspace{0.1cm}
\doublerulesep 1pt \tabcolsep 0.08in
\small{
\begin{tabular}{cc|c|c|c|c|c|}  \cline{3-7}
& & $ \mu \pi^+ \pi^-$       &     $\mu \rho$     & $\mu f_0 $ &   $3 \mu$    & $\mu \gamma$   \\   \cline{1-7}
\multicolumn{1}{ |c| }{\multirow{2}{*}{  $\mathrm{D}$ } }  &  $R_{F,D}$  &  $0.26 \times10^{-2}$  & $0.22 \times10^{-2}$   &  \eval{0.13}{3} & \eval{0.22}{2} & 1  \\  %\cline{2-7}
%  \rowcolor{Gray}
 \multicolumn{1}{ |c| }{}   &   $\mathrm{BR}$ &  $< 1.1 \times 10^{-10}$  & $<  9.7 \times 10^{-11}$   &  $< 5.7 \times 10^{-12}$ & $< 9.7 \times 10^{-11}$  & $< 4.4 \times10^{-8} $  \\ \cline{1-7}
 \multicolumn{1}{ |c| }{\multirow{2}{*}{$ \mathrm{S}  $ } }  &  $R_{F,S}$  &  1  &  $0.28$  &  $0.7$ &   - &  - \\  %\cline{2-7}
%   \rowcolor{Gray}
 \multicolumn{1}{ |c| }{}   &   $\mathrm{BR}$ & $<$ \eval{2.1}{8}          & $<$  \eval{5.9}{9}  & $<$ \eval{1.47}{8} &-  &  - \\  \cline{1-7}
 \multicolumn{1}{ |c| }{\multirow{2}{*}{$ \mathrm{V^{(\gamma)} } $ } }  &  $R_{F,V^{(\gamma)}}$  &  1  &  $0.86$  &  $0.1$ &   - &  - \\  %\cline{2-7}
%   \rowcolor{Gray}
 \multicolumn{1}{ |c| }{}   &   $\mathrm{BR}$ & $<$ \eval{1.4}{8}          & $<$  \eval{1.2}{8}  & $<$ \eval{1.4}{9} &-  &  - \\ \cline{1-7}
  \multicolumn{1}{ |c| }{\multirow{2}{*}{$ \mathrm{Z} $ } }  &  $R_{F,Z}$  &  1  &  $0.86$  &  $0.1$ &   - &  - \\  %\cline{2-7}
%   \rowcolor{Gray}
 \multicolumn{1}{ |c| }{}   &   $\mathrm{BR}$ & $<$  \eval{1.4}{8}          & $<$  \eval{1.2}{8}  & $<$ \eval{1.4}{9} &-  &  - \\ \cline{1-7}
  \multicolumn{1}{ |c| }{\multirow{2}{*}{$\mathrm{G}$ } }  &  $R_{F,G}$  &  1  &  $0.41$  &  $0.41$ &   - &  - \\  %\cline{2-7}
%   \rowcolor{Gray}
 \multicolumn{1}{ |c| }{}   &   $\mathrm{BR}$ & $<$ \eval{2.1}{8}          & $<$  \eval{8.6}{9}  & $<$ \eval{8.6}{9} &-  &  - \\ 
\hline\hline
\end{tabular}}
\end{center}
\end{table}
%%%%%%%%%%%%%%%%%%%%%%%%%%%%%%%%%%%%%%%%%%%%%%

%%%%%%%%%%%%%%%%%%%%%%%%%%%%%%%%%%%%%%%%%%%%%%%%%%%%%%%%%%%%%%%%%%%%%%%%%%
\begin{table}[t]
\setlength{\extrarowheight}{3pt}
\begin{center}
\caption{\label{tab::patternsI} \it \small  Expected pattern for the branching ratio of various semileptonic $\tau \rightarrow \mu P$ decays within the single operator dominance hypothesis.    }
\vspace{0.1cm}
\doublerulesep 1pt \tabcolsep 0.13in
\small{
\begin{tabular}{cc|c|c|c|}  \cline{3-5}
 &   & $  \mu \pi$       &     $\mu \eta$  & $\mu \eta^{\prime}$      \\ \cline{1-5}
 \multicolumn{1}{ |c| }{\multirow{2}{*}{ $\mathrm{Z}$  } }  &  $R_{F,Z}$  & 1  &   0.3    &   0.28  \\  
%  \rowcolor{Gray}
 \multicolumn{1}{ |c| }{   }   &   $\mathrm{BR}$ &  $<$ \eval{1.1}{7}    & $<  3.3 \times 10^{-8}$  & $<$ \eval{3.1}{8}      \\ \cline{1-5}
\multicolumn{1}{ |c| }{\multirow{2}{*}{ $\mathrm{\widetilde G}$  } }  &  $R_{F,\widetilde{G}}$  & -   &   0.25    &   1  \\   
%  \rowcolor{Gray}
 \multicolumn{1}{ |c| }{   }   &   $\mathrm{BR}$ & -  & $<  3.25 \times 10^{-8}$  & $<$ \eval{1.3}{7}      \\ \cline{1-5}
 \multicolumn{1}{ |c| }{\multirow{2}{*}{  $\mathrm{ P^{(1)   }   }$ } }  &  $R_{F,P^{(1)}}$  &  -  &    0.97  & 1  \\
%   \rowcolor{Gray}
 \multicolumn{1}{ |c| }{}   &   $\mathrm{BR}$ & -  &    $<$  \eval{6.5}{8}    & $<$  \eval{6.7}{8}            \\ \cline{1-5}
 \multicolumn{1}{ |c| }{\multirow{2}{*}{  $\mathrm{ P^{(2)   }   }$ } }  &  $R_{F,P^{(2)}}$  &  0.005   & 1  &   0.94  \\
%   \rowcolor{Gray}
 \multicolumn{1}{ |c| }{}   &   $\mathrm{BR}$ &  $<$ \eval{3.25}{10}    & $<$ \eval{6.5}{8}    & $<$ \eval{6.1}{8}     \\ 
\hline\hline
\end{tabular}}
\end{center}
\end{table}
%%%%%%%%%%%%%%%%%%%%%%%%%%%%%%%%%%%%%%%%%%%%%%

%
%
%
\subsection{The discriminating power of differential distributions}

The discriminating power of differential distributions in many-body decays to different kinds of NP is well known in flavor physics.
The limiting factor for these kind of analyses for LFV $\tau$ decays is clear.  Assuming that some of these transitions are within reach of Belle II and are observed at some point, the expected number of events that can be gathered in the near future will be very low.    Without being pessimistic, just the observation of LFV in the charged lepton sector would constitute an indisputable signal of physics beyond the SM and would certainly motivate further efforts to understand its origin.    In this sense, extracting information from the differential distributions in three-body LFV $\tau$ decays seems a straightforward goal if these transitions are observed in the future.  Together with correlations between the BR of different LFV $\tau$ decay channels, differential distributions are probably the most accessible way to gain information about the underlying dynamics responsible for LFV in $\tau$ decays.      Other possibilities would be to study observables involving polarized $\tau$ decays~\cite{Matsuzaki:2007hh,Giffels:2008ar} or searches for $\mu N \rightarrow \tau X$ conversion with high-intensity and high-energy muon beams~\cite{Sher:2003vi,Kanemura:2004jt}, though we will not explore this here.

\subsubsection{The semileptonic decay $\tau \rightarrow  \mu \pi^+ \pi^-$}

The invariant mass of the pion pair in $\tau \rightarrow \mu \pi^+ \pi^-$ decays contains information about the underlying NP responsible for LFV.   The crucial point to extract reliable results is a proper determination of the relevant hadronic form factors in all the kinematical range available to the $\pi \pi$ pair.    Recent progress in the determination of the hadronic form factors for $\tau \rightarrow \mu \pi^+ \pi^-$ decays  has  been achieved in Refs.~\cite{Daub:2012mu,Celis:2013xja}, improving considerably over previous treatments in the literature.
A brief discussion of the needed form factors is given in Appendix~\ref{hadronic}.  

\begin{figure}[ht!]
\centering
\includegraphics[width=0.5\textwidth]{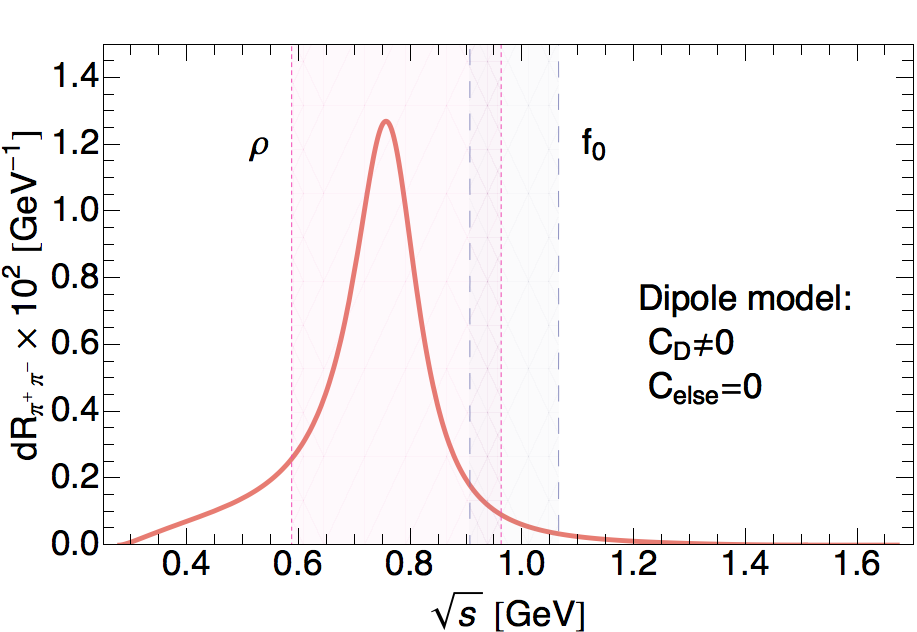}
\caption{\label{fig:Dipolepipi} \it \small   Differential ratio $dR_{\pi^+\pi^-}$ as a function of the pion invariant mass spectrum in $\tau \rightarrow \mu \pi^+ \pi^-$ decays, assuming dipole operator dominance.   Experimental cuts on the pion invariant mass used to set limits on $\tau \rightarrow \mu \rho$ and $\tau \rightarrow \mu f_0$ decays are shown as pink (short-dashed borders) and gray (long-dashed borders) bands respectively. }
\end{figure}

\begin{figure}[ht!]
\centering
\includegraphics[width=0.45\textwidth]{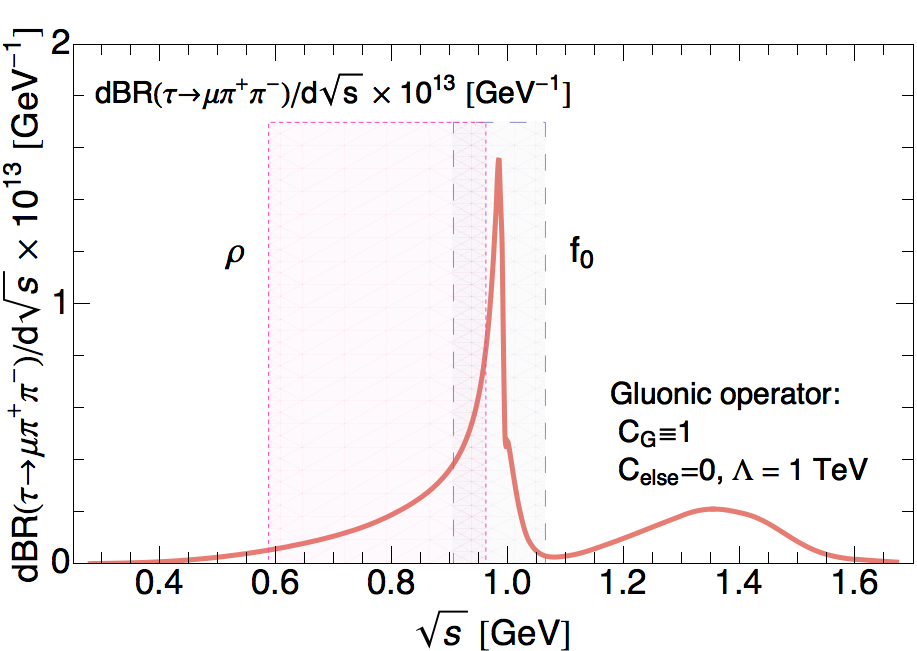}
~
\includegraphics[width=0.45\textwidth]{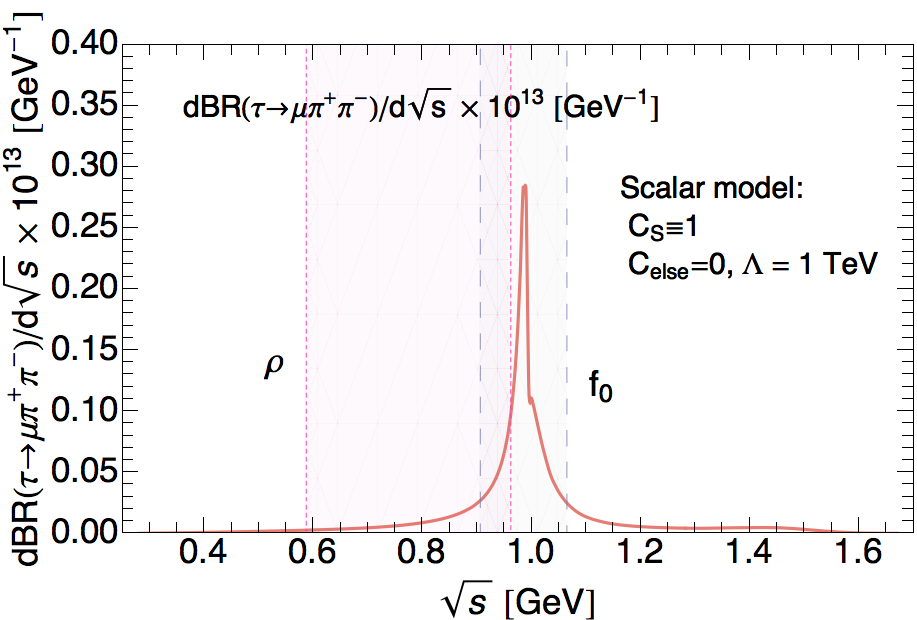}
\caption{\label{fig:gluonicpipi} \it \small   Differential branching ratio, $d\mathrm{BR}(\tau \rightarrow \mu \pi^+ \pi^-)/d\sqrt{s}$, as a function of the pion invariant mass spectrum in the Gluonic model (left) and in the Scalar model (right).   Colored bands are defined as in Fig.~\ref{fig:Dipolepipi}.  }
\end{figure}

In the Dipole model the pion invariant mass spectrum is determined by the pion vector form factor and peaks around the $\rho$ mass.    In Fig.~\ref{fig:Dipolepipi} we plot the ratio
\be
dR_{\pi^+\pi^-} \equiv \frac{d\Gamma(\tau \rightarrow \mu \pi^+ \pi^-  )/d\sqrt{s}}{ \Gamma(\tau \rightarrow \mu \gamma) } \,,
\ee
for  the Dipole model.   Note that in this case all the dependence on $\mathrm{C_D}/\Lambda^2$ cancels in this ratio.   The decays $\tau \rightarrow \mu \rho$ and $\tau \rightarrow \mu f_0$ are measured by applying a cut on the invariant mass of the pion pair, the corresponding intervals are shown as pink (short-dashed borders) and gray (long-dashed borders) bands in Fig.~\ref{fig:Dipolepipi}.     Both in the Vector and $Z$-penguin models the invariant mass spectrum is also determined by the pion vector form factor, so it has the same form than in the Dipole model.      The Scalar and Gluonic models involve new form factors: in this case the pion invariant mass spectrum peaks around the $f_0(980)$ resonance as shown in Fig.~\ref{fig:gluonicpipi}.    
In the Gluonic model a long tail is produced towards low invariant pion masses and a secondary peak appears around $\sqrt{s} \sim 1.4$~GeV, due to the $f_0(1370)$ and $f_0(1500)$.  
In the Scalar model these features are less pronounced. 
 
%%%%%%%%%%%%%%%%%%%%%%%%%%%%%%%%%%%%%%%%%%%%%%%%%%%%%%%%%%%%%%%%%%%%%%%%%%
%%%%%%%%%%%%%%%%%%%%%%%%%%
%%%%%%%

%%%%%%%%%%%%%%%%%%%%%%%%%%%%%%%%%%%%%%%%%%%%%%%%%%%%%%%%%%%%%%%%%%%%%%%%%%
%%%%%%%%%%%%%%%%%%%%%%%%%%
%%%%%%%

%%%%%%%%%%%%%%%%%%%%%%%%%%%%%%%%%%%%%%%%%%%%%%%%%%%%%%%%%%%%%%%%%%%%%%%%%%

\subsubsection{The leptonic decay $\tau \rightarrow 3 \mu$} \label{sst3mu}

\begin{figure}[ht!]
\centering
\includegraphics[width=0.47\textwidth]{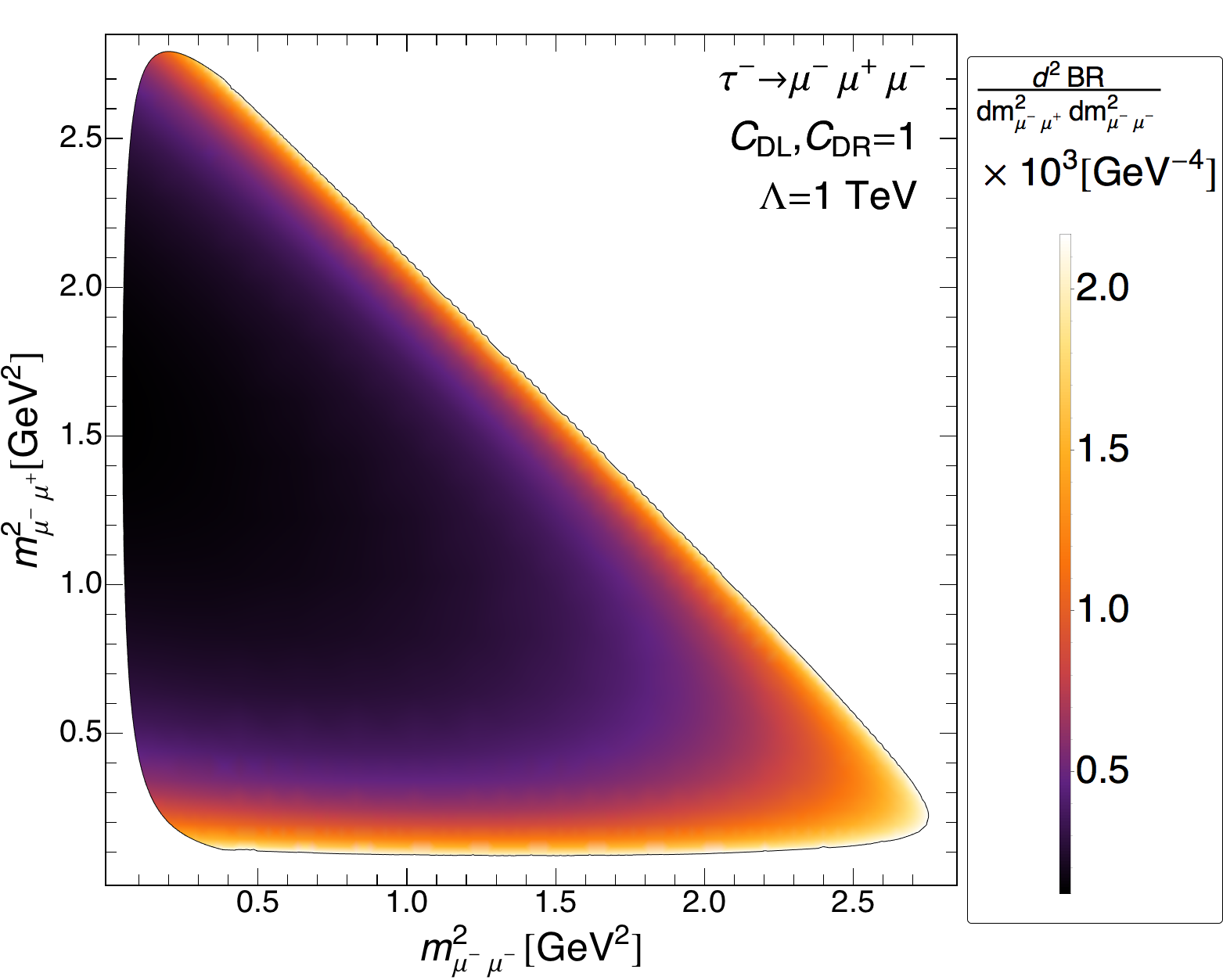}
~~~
\includegraphics[width=0.47\textwidth]{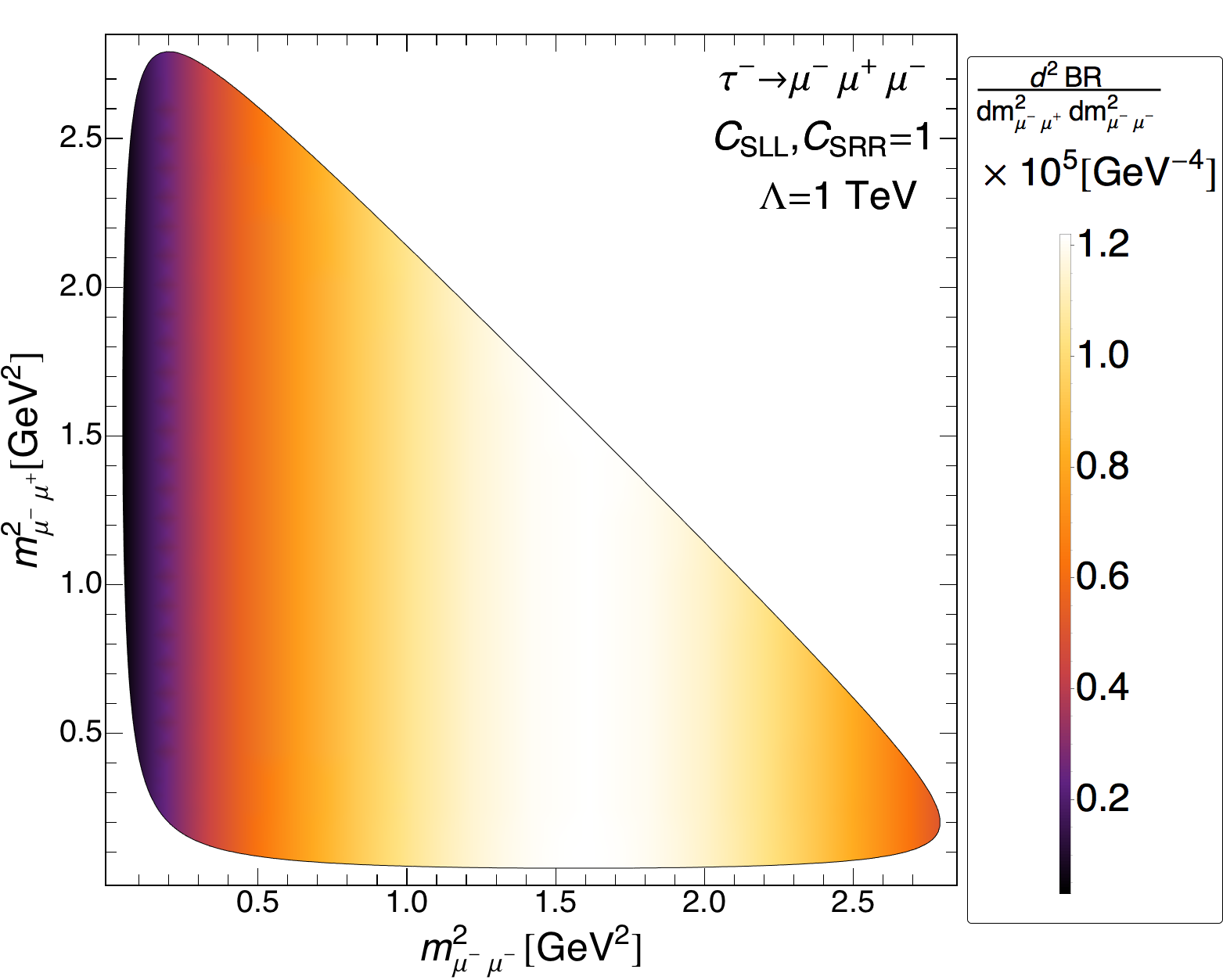}
\caption{\label{fig:Dalitz2} \it \small Dalitz plot for $\tau^- \rightarrow \mu^- \mu^+ \mu^-$ decays when all operators are assumed to vanish with the exception of $\mathrm{C_{DL,DR}} = 1$ (left) and $\mathrm{C_{SLL,SRR}} = 1$ (right), taking $\Lambda = 1$~TeV in both cases.  Colors denote the density for $d^2 \mathrm{BR}/(  dm_{\mu^-\mu^+}^2 dm_{\mu^-\mu^-}^2  )$, small values being represented by darker colors  and large values in lighter ones.  
Here $m_{\mu^-\mu^+}^2$ represents $m_{12}^2$ or $m_{23}^2$, defined in Sec.~\ref{sect:rad}.
 }
\end{figure}

\begin{figure}[ht!]
\centering
\includegraphics[width=0.47\textwidth]{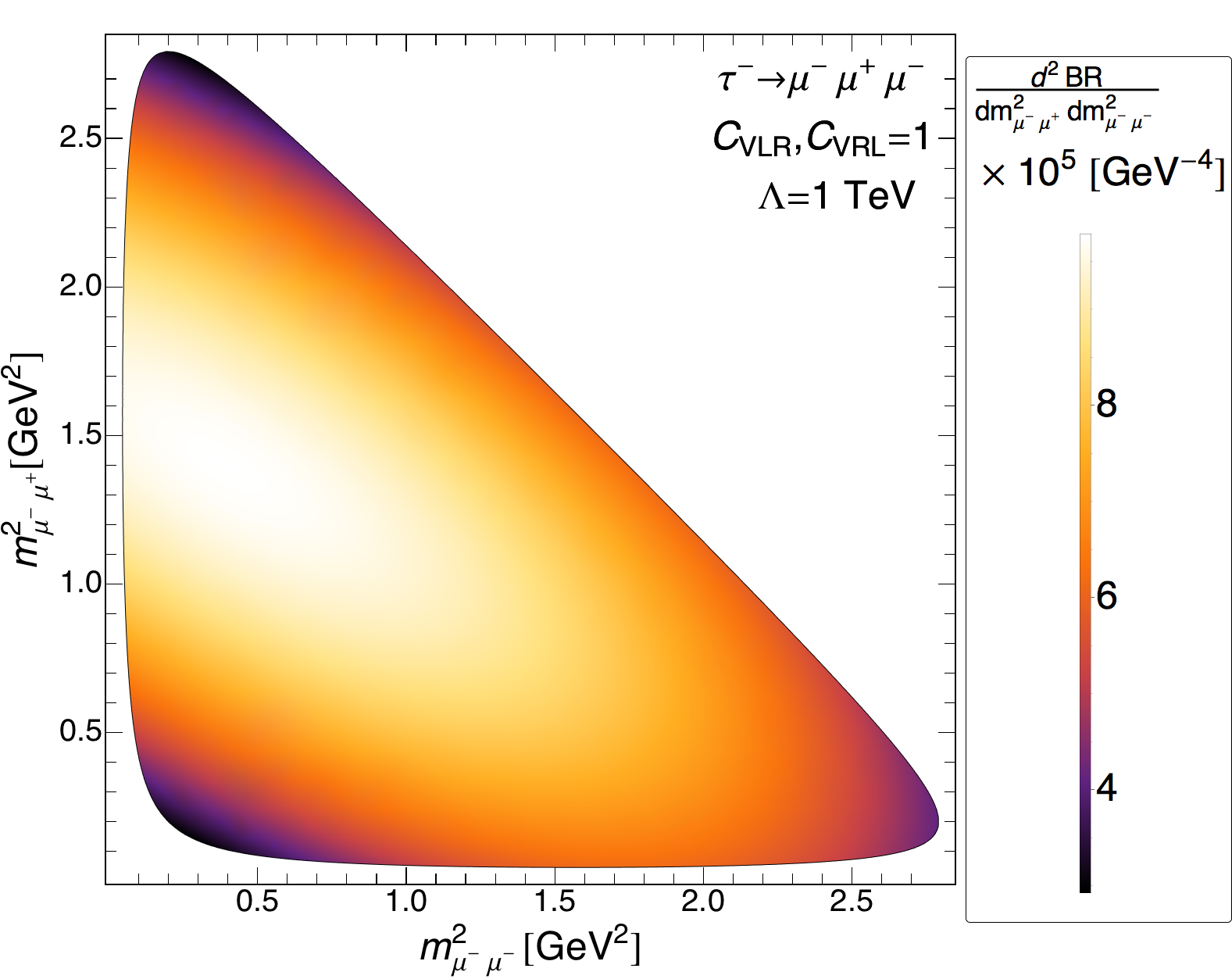}
~~~
\includegraphics[width=0.47\textwidth]{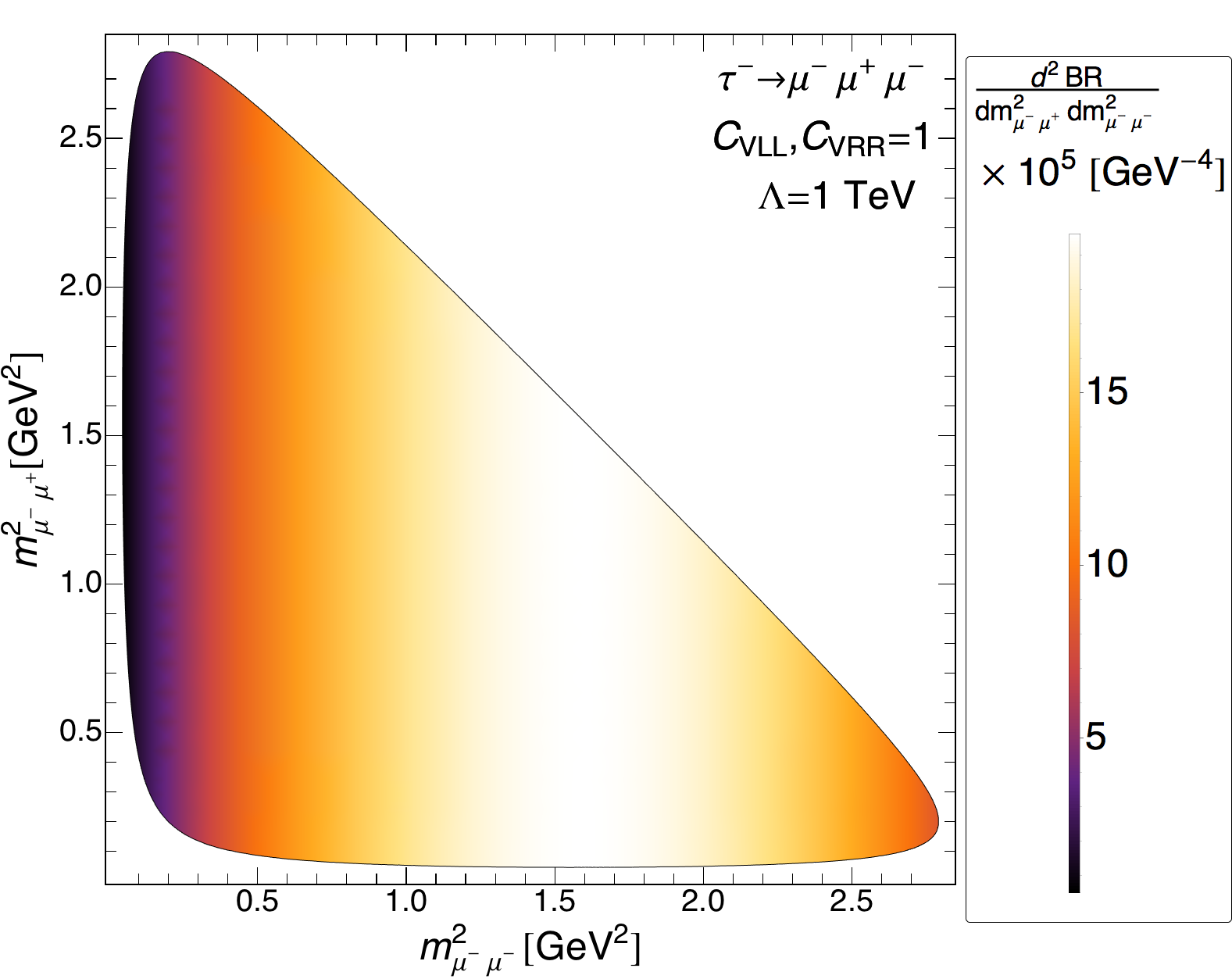}
\caption{\label{fig:Dalitz} \it \small Dalitz plot for $\tau^- \rightarrow \mu^- \mu^+ \mu^-$ decays when all operators are assumed to vanish with the exception of $\mathrm{C_{VRL,VLR}} = 1$ (left) and $\mathrm{C_{VLL,VRR}} = 1$ (right), taking $\Lambda = 1$~TeV in both cases.  Colors are defined as in Fig.~\ref{fig:Dalitz2}.  }
\end{figure}

\begin{figure}[ht!]
\centering
\includegraphics[width=0.5\textwidth]{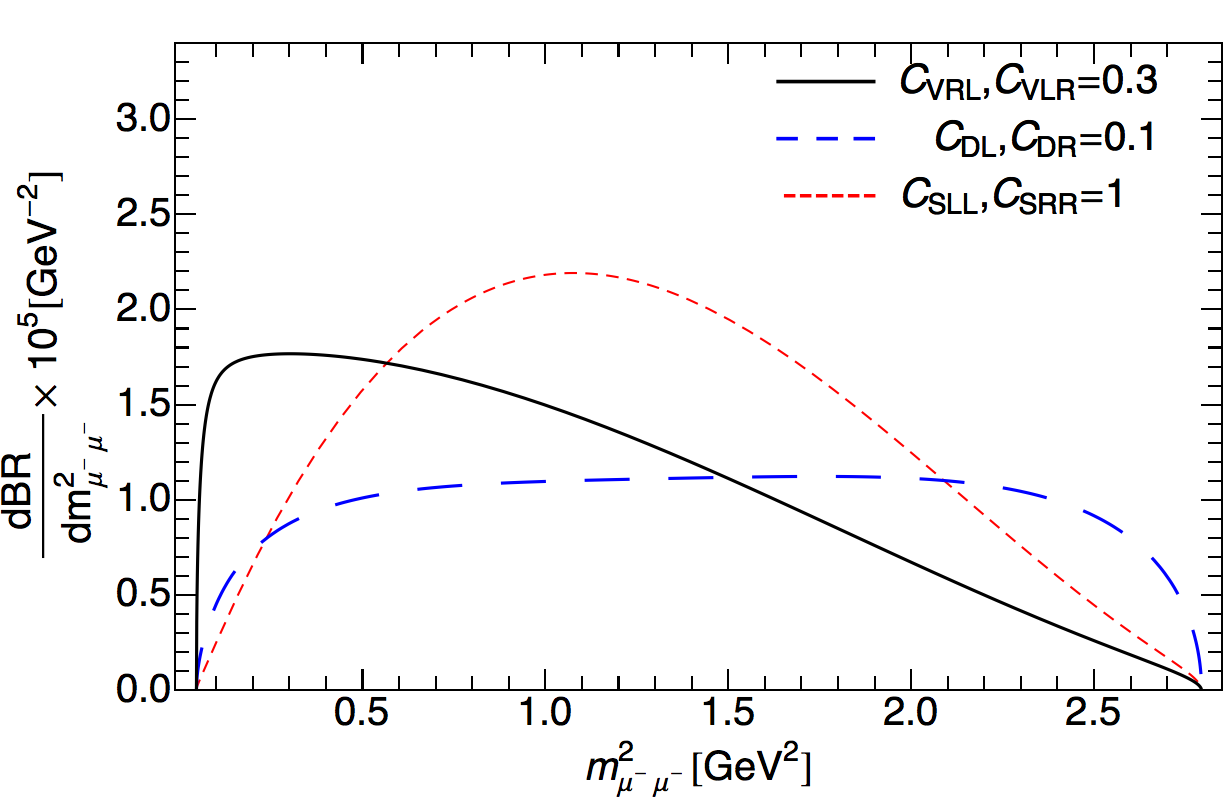}
\caption{\label{fig:specm13s} \it \small  Same sign di-muon invariant mass spectrum for $\tau^- \rightarrow \mu^- \mu^+ \mu^-$ decays when all operators are assumed to vanish with the exception of $\mathrm{C_{VLR, VRL}} = 0.3$ (continuous black), $\mathrm{C_{DL, DR}} = 0.1$ (long-dashed blue) and $\mathrm{C_{SLL, SRR}} = 1$ (short-dashed red), taking $\Lambda = 1$~TeV. }
\end{figure}

Differential distributions for three-body decays also provide valuable information in the case of LFV leptonic decays.   A Dalitz plot analysis of $\tau^- \rightarrow  \mu^- \mu^+ \mu^-$ decays for example can be used to discriminate among different effective operators. 
In the case where dipole operators dominate, the distribution of events in the Dalitz plot concentrates on borders of the phase space as shown in Fig.~\ref{fig:Dalitz2} (left-plot).\footnote{We have kept the muon mass at its physical value for obtaining Figs.~\ref{fig:Dalitz2}, \ref{fig:Dalitz} and \ref{fig:specm13s}.}  Other effective operators also produce distinctive patterns on a Dalitz plot, see Figs.~\ref{fig:Dalitz2} and \ref{fig:Dalitz}.    One would expect a flat distribution for the same-sign muon invariant mass spectrum $(d \mathrm{BR}  /dm_{\mu^-\mu^-}^2)$  
in the case of dipole operators as shown in Fig.~\ref{fig:specm13s}.   The vector operators $\mathrm{C_{VRL,VLR}}$ would produce a spectrum peaked towards low invariant masses $m_{\mu^- \mu^-}^2$, the scalar operators  $\mathrm{C_{SLL, SRR}}$ on the other hand would give rise to a peaked spectrum around $m_{\mu^- \mu^-}^2 \sim 1$~GeV$^2$, see Fig.~\ref{fig:specm13s}.  The discrimination of different kinds of NP through a Dalitz plot analysis in LFV leptonic $\tau$ decays has been discussed in detail in Refs.~\cite{Dassinger:2007ru,Matsuzaki:2007hh}.

%%%%%%%%%%%%%%%%%%%%%%%%%%%%%%%%%%%%%%%%%%%%%%%%%%%%%%%%%%%%%%%%%%%%%%%%%%
%%%%%%%%%%%%%%%%%%%%%%%%%%
%%%%%%%

%%%%%%%%%%%%%%%%%%%%%%%%%%%%%%%%%%%%%%%%%%%%%%%%%%%%%%%%%%%%%%%%%%%%%%%%%%
%%%%%%%%%%%%%%%%%%%%%%%%%%
%%%%%%%

   %%%%%%%%%%%%%%%%%%%%%%%%%%%%%%%%%%%%%%%%%%%%%%%%%%%%%%%%%%%%%%%%%%%%%%%%%%
%%%%%%%%%%%%%%%%%%%%%%%%%%
%%%%%%%

%
%
%
 %
\section{Future prospects}
\label{sec:future}
Present experimental limits on LFV $\tau$ decays are at the $10^{-8}$ level thanks to the large amount of data collected at Belle and BaBar.  As a comparison,  before Belle and BaBar the best upper bound on $\mathrm{BR}(\tau \rightarrow \mu \gamma)$ was set at the CLEO detector with $L \sim13.8~\text{fb}^{-1}$ of integrated luminosity, finding $\mathrm{BR}(\tau \rightarrow \mu \gamma) < 1.1 \times 10^{-6}$ ($90\%$~CL)~\cite{Ahmed:1999gh}.    Belle and BaBar have finally stopped collecting data, reaching a final integrated luminosity of $L \gtrsim1~\text{ab}^{-1}$ and $L \sim 550~\text{fb}^{-1}$ respectively.   The upcoming Belle II experiment at the SuperKEKB collider is expected to deliver $L \sim 50$~ab$^{-1}$ of 
data~\cite{Abe:2010sj}.    In cases where the number of background events is not negligible, the $90\%$~CL upper limit on the BR $(\mathrm{BR}_{90})$ is expected to improve with the integrated luminosity $L$ as $\mathrm{BR}_{90} \propto 1/\sqrt{L}$.  One can then expect an improvement of the present upper bounds by a factor of ten approximately with $L \sim 50$~ab$^{-1}$ of collected data at Belle II.     Prospects for LFV $\tau$ decays at a Super Tau-Charm Factory are also encouraging, with an estimated sensitivity of $\mathrm{BR}(\tau \rightarrow \mu \gamma) \lesssim 10^{-9}$ with $10$~ab$^{-1}$~\cite{Bobrov:2012kg}.    

\begin{figure}[ht!]
\centering
\includegraphics[width=0.7\textwidth]{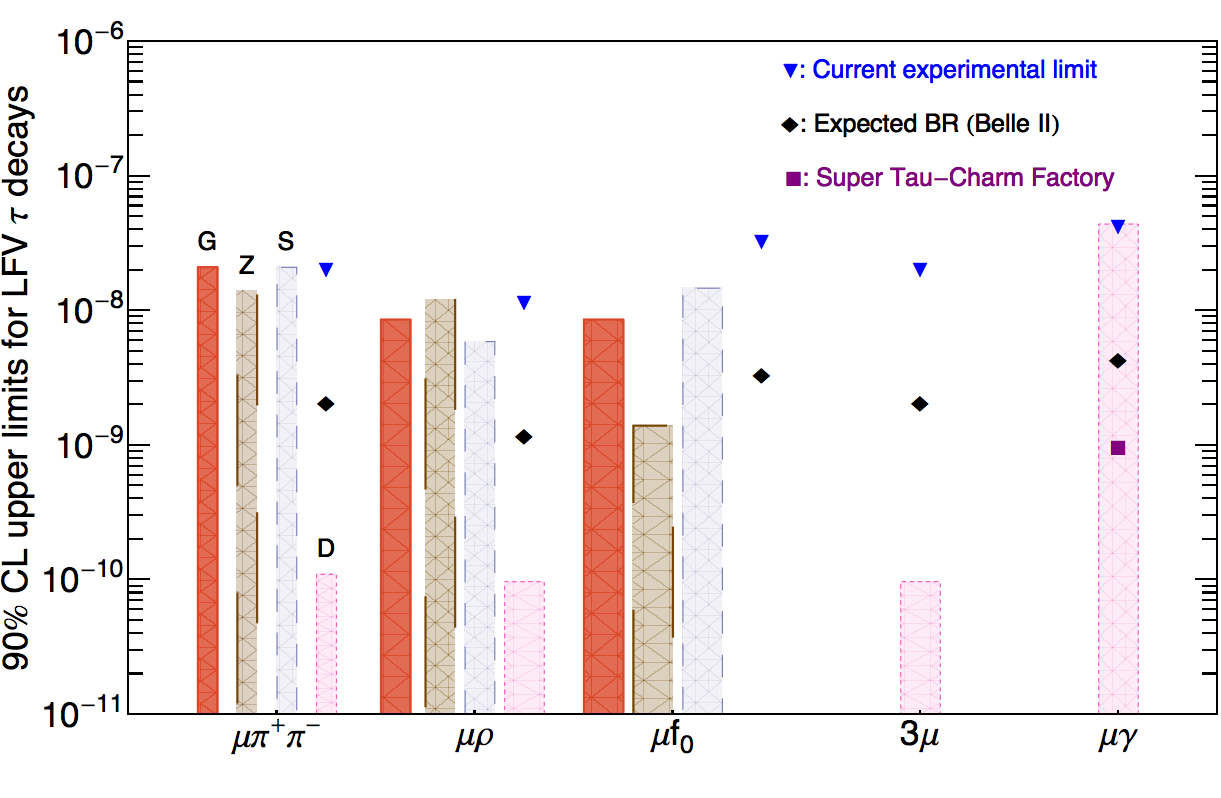}
\caption{\label{fig:prospects} \it \small  
Prospects for the observation of LFV $\tau$ decays.   Current experimental limits on the BRs at $90\%$~CL are given as well as expected limits at future machines.     Vertical bars represent bounds on the BRs derived from the non-observation of LFV $\tau$ decays in the different benchmark models for single operator dominance.          }
\end{figure}

\begin{figure}[ht!]
\centering
\includegraphics[width=0.6\textwidth]{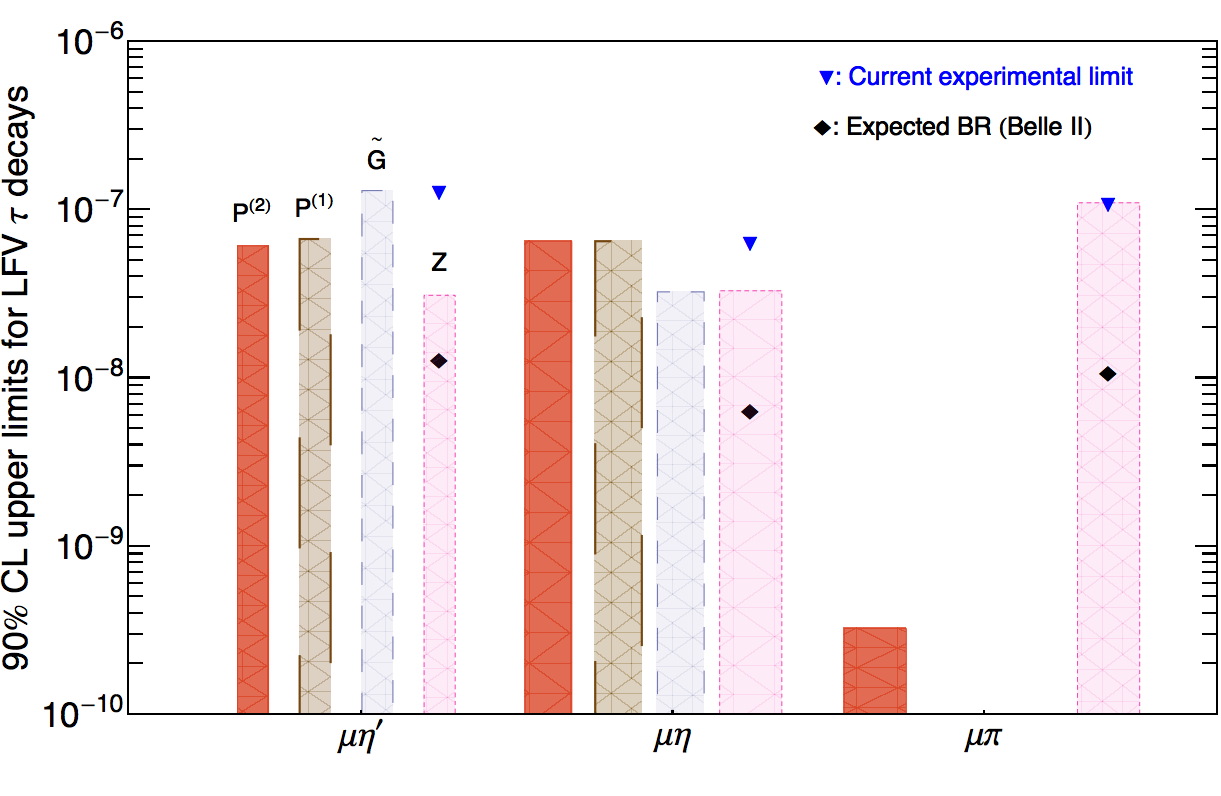}
\caption{\label{fig:prospectsII} \it \small  Prospects for the observation of LFV $\tau \rightarrow \mu P$ decays.   Other captions are the same than for Fig.~\ref{fig:prospects}.   }
\end{figure}

 In Figs.~\ref{fig:prospects} and \ref{fig:prospectsII} we show future prospects for the observation of LFV $\tau$ decays.   The figures show (i) current experimental upper limits on the BRs at $90\%$~CL;
(ii) expected future limits assuming an improvement of the sensitivity by a factor of ten;     
(iii) upper bounds  (colored bars) that can be derived on the BRs, within each of the benchmark models for single operator dominance, 
from the non-observation of LFV $\tau$ decays (from Section~\ref{phenosec}). 
Among other features,  Fig.~\ref{fig:prospects} implies that if the dipole operator dominates, clearly 
$\tau \to \mu \gamma$ is the channel to focus on (the other have limits below future sensitivity). 
However, if other operators contribute, then hadronic decays offer greater discovery potential, so they should be 
vigorously pursued.

   %%%%%%%%%%%%%%%%%%%%%%%%%%%%%%%%%%%%%%%%%%%%%%%%%%%%%%%%%%%%%%%%%%%%%%%%%%
%%%%%%%%%%%%%%%%%%%%%%%%%%
%%%%%%%

%%%%%%%%%%%%%%%%%%%%%%%%%%%%%%%%%%%%%%%%%%%%%%%%%%%%%%%%%%%%%%%%%%%%%%%%%%
%%%%%%%%%%%%%%%%%%%%%%%%%%
%%%%%%%

\begin{figure}[ht!]
\centering
\includegraphics[width=0.45\textwidth]{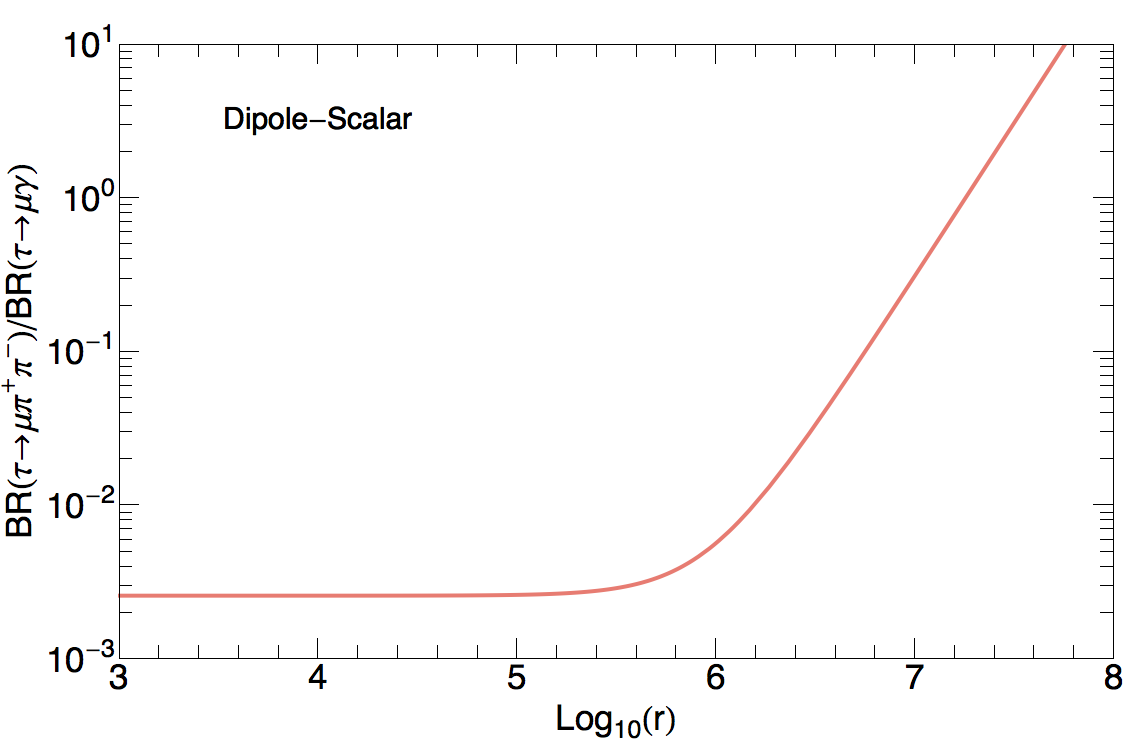} \\
\includegraphics[width=0.4\textwidth]{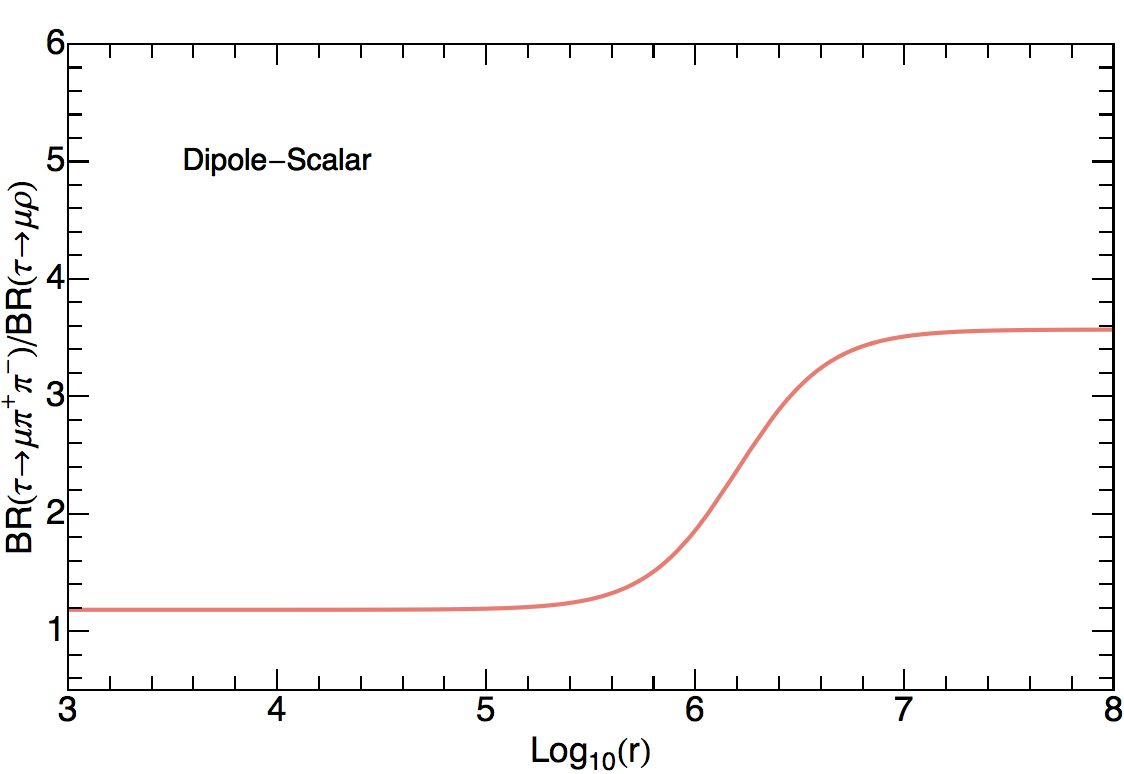}
~
\includegraphics[width=0.4\textwidth]{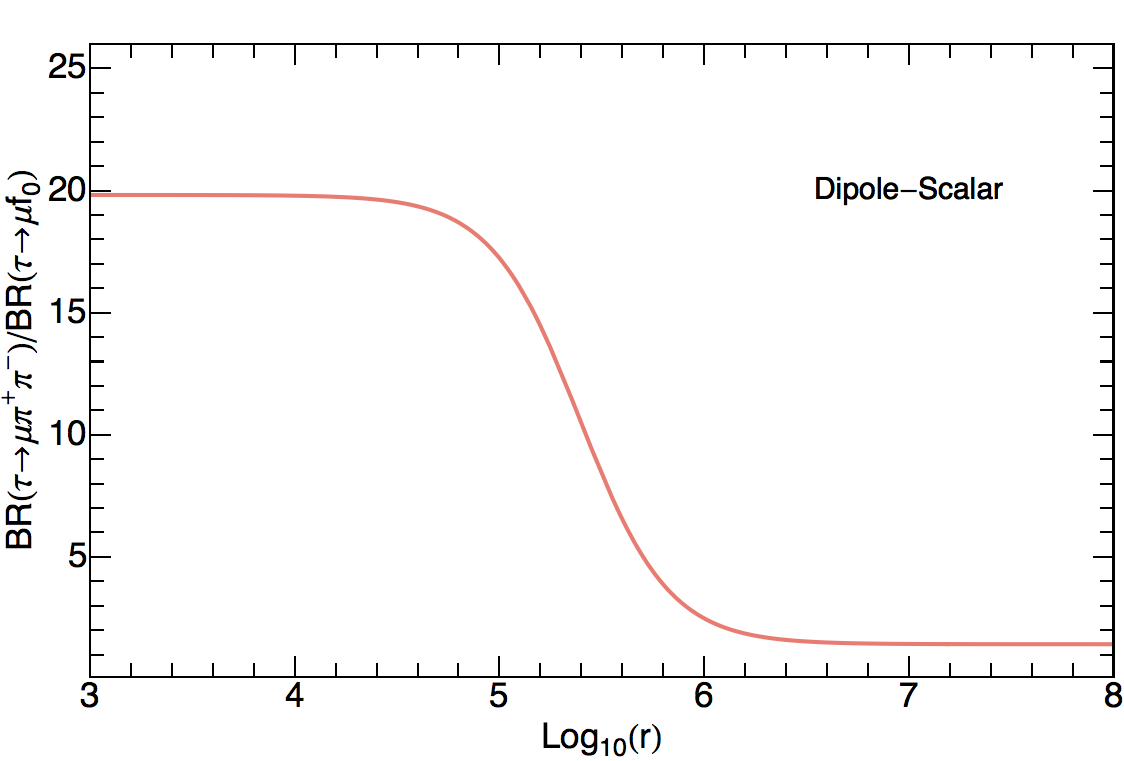}
\caption{\label{fig:DipoleScalar} \it \small   Dipole-Scalar model:    Ratios $\mathrm{BR}( \tau \rightarrow \mu \pi^+ \pi^-)/\mathrm{BR}( \tau \rightarrow \mu \gamma )$ (top), $\mathrm{BR}( \tau \rightarrow \mu \pi^+ \pi^-)/\mathrm{BR}( \tau \rightarrow \mu \rho )$ (bottom-left), and $\mathrm{BR}( \tau \rightarrow \mu \pi^+ \pi^-)/\mathrm{BR}( \tau \rightarrow \mu f_0 )$ (bottom-right) as a function of $\mathrm{Log}_{10}(r)$, with $r = |\mathrm{C_S}/\mathrm{C_D}|$.}
\end{figure}

So far we have discussed the implications of the single operator dominance hypothesis in a series of benchmark scenarios.  Due to operator mixing under the renormalization group evolution, one would actually expect that several operators are relevant at the low energy scale.     To analyze this situation, let us  consider a simple example.  We define the Dipole-Scalar model in which both dipole and scalar operators are present at the same time,

\be
 \mathrm{C_D} \equiv  \mathrm{C_{DL}}   \neq 0 \,, \qquad  \mathrm{ C_S \equiv  C_{SL}^{u}  =  C_{SL}^{d}  = C_{SL}^{s}  }   \neq 0 \,, \qquad \mathrm{C_{else}} = 0     \,.
\ee
In this case only the parameters $\mathrm{C_{D}}/\Lambda^2$ and $r \equiv |\mathrm{C_S}/ \mathrm{C_D}|$ appear.    The radiative $\tau \rightarrow \mu \gamma$ and semileptonic decays $\tau \rightarrow \mu \pi^+ \pi^-,  \mu \rho, \mu f_0$ receive contributions at tree-level from these operators.   Note from Eq.~\eqref{eqforpipi} that there is no interference between dipole and scalar contributions so that there is no sensitivity to the sign of $\mathrm{C_S}/ \mathrm{C_D}$.  It is possible to test the two-operator dominance hypothesis by 
(i) taking ratios of the BRs in the different decay modes (see Fig.~\ref{fig:DipoleScalar}), and also by (ii) analyzing the $\pi \pi$ spectrum in 
$\tau \to \mu \pi^+ \pi^-$, where both the $\rho$ and $f_0$ features will appear, with relative strength controlled by the ratio of Wilson coefficients. 
An explicit example of this is given by non-standard LFV Higgs couplings, that generate both  dipole and scalar operators. 
The resulting spectrum is shown in Ref.~\cite{Celis:2013xja}.

%%%%%%%%%%%%%%%%%%%%%%%%%%%%%%%%%%%%%%%%%%%%%%%%%%%%%%%%%%%%%%%%%%%%%%%%%%
%%%%%%%%%%%%%%%%%%%%%%%%%%
%%%%%%%

%
%
%
%
%
\section{Conclusions}
\label{sum}

In this work we have analyzed the model-discriminating power of lepton flavor violating $\tau$ decays within an effective field theory framework, including  radiative, purely leptonic, and semileptonic decay modes.
The vast majority of  available phenomenological studies has focused on the radiative and leptonic LFV $\tau$ decays, in part because these do not suffer from the hadronic uncertainties present in semileptonic $\tau$ decays 
and also because many NP scenarios predict large rates for these modes.  One has to keep in mind,  however,  that these decays are only sensitive to particular operators which might be  suppressed in some NP models or
for some regions of the NP parameter space.   As illustrated in Table~\ref{tab::sensitivityP},  semileptonic $\tau$ decays are then complementary modes in our search for LFV in charged leptons, in that they probe a larger set of operators.
 
Compared with previous discussions in the literature, our main contribution is  that we incorporate in our analysis recent developments on the determination of the hadronic form factors for $\tau \rightarrow \ell \pi \pi$ ($\ell = e, \mu$) decays~\cite{Daub:2012mu,Celis:2013xja}.   Previous treatments of the form factors based on ChPT fail to describe properly the hadronic dynamics because the invariant mass of the pion pair $\sqrt{s} \leq (m_{\tau} - m_{\ell})$ can be well outside the range of validity of ChPT.     
A proper determination of the hadronic matrix elements in $\tau \rightarrow \ell \pi \pi$ decays is crucial not only  to  obtain a reliable estimate of the decay rate and meaningful bounds on the NP parameters, but also to extract information about the underlying dynamics responsible for LFV from the  pion pair  invariant mass distribution. 
(The interpretation of $\tau \rightarrow \ell \rho$ and $\tau \rightarrow \ell f_0$ searches within NP models also requires a correct description of the hadronic matrix elements as implemented here.)

LFV  $\tau$ decays offer two main  handles to  discriminate among underlying models of new physics, i.e. to identify which operators are present at low energy and  what 
is their relative strength: 
\begin{itemize}
\item The first handle is provided by correlations among the different LFV $\tau$ decay rates.
To illustrate this, after defining several benchmark scenarios in which only one type of operator dominates, 
in Tables \ref{tab::patternsII} and \ref{tab::patternsI}  we presented   the pattern of LFV branching ratios for each benchmark model. 

\item The second handle is provided by  differential distributions in many-body decays, such as $\tau \to \mu \pi^+ \pi^-$ and $\tau \to 3 \mu$.
We showed   how  the  analysis of the two-pion invariant mass spectrum in $\tau \rightarrow \ell \pi \pi$  decays can be used to disentangle different effective operators (see Figs.~\ref{fig:Dipolepipi} and  \ref{fig:gluonicpipi}). 
We also discussed the discrimination of different operators contributing to leptonic $\tau \rightarrow 3 \mu$ decays based on a Dalitz plot analysis (see Figs.~\ref{fig:Dalitz2}, \ref{fig:Dalitz}, and \ref{fig:specm13s}): our results in this respect are very similar to those presented previously in Ref.\cite{Dassinger:2007ru}.

\end{itemize}

We have also examined future prospects for the observation of LFV $\tau$ decays, discussing the discovery potential of each decay mode within the various benchmark models (see Figs.~\ref{fig:prospects} and \ref{fig:prospectsII}).
 Our results imply that $\tau \to \mu \gamma$ is the most promising  channel only if the dipole operator dominates: in this scenario the other  modes  have branching ratios  below future sensitivity. 
On the other hand, in new physics models in which the dipole operator is not the dominant one, semileptonic  decays  such as $\tau \to \ell  \pi \pi$ ($\ell = e, \mu$)  offer greater discovery potential, so they should be 
definitely  pursued in order to maximize the impact of future flavor factories.

\begin{appendix}
\section{Hadronic matrix elements}
\label{hadronic}
In this appendix we provide a brief discussion of the relevant hadronic matrix elements needed to describe the different semileptonic $\tau$ decays considered in this work.

\subsection{Semileptonic $\tau \rightarrow \mu P$ decays}
 The relevant hadronic matrix elements for the evaluation of semileptonic $\tau \rightarrow \mu P$ decays (where $P$ is a pseudoscalar meson) can be obtained following the Feldmann-Kroll-Stech (FKS) mixing scheme~\cite{Feldmann:1998vh}, reviewed in Ref.~\cite{Beneke:2002jn}.     Pseudoscalar and axial current densities are parametrized in terms of a series of parameters which encode the non-perturbative QCD dynamics:
\begin{align}  \label{mesonII}
 \langle \pi^0(p)  |  \bar u  \, \gamma_5 \, u   |  0 \rangle &\;=\;    i \, \frac{m_{\pi}^2}{2 \sqrt{2} \hat m} \, f_{\pi}  \,,  \qquad  \qquad  \langle \pi^0(p)  |  \bar d \, \gamma_5  \,d   | 0 \rangle = - \langle \pi^0(p)  |  \bar u  \, \gamma_5 \, u   |  0 \rangle  \,,  \nonumber   \\ 
 \langle \pi^0(p)  |  \bar u  \, \gamma^{\mu} \gamma_5 \, u   |  0 \rangle &\;=\;    \frac{i}{\sqrt{2}}   \,    f_{\pi} \,  p^{\mu}  \,,  \qquad  \qquad  \langle \pi^0(p)  |  \bar d \,  \gamma^{\mu} \gamma_5  \,d   | 0 \rangle = - \langle \pi^0(p)  |  \bar u  \, \gamma^{\mu} \gamma_5 \, u   |  0 \rangle   \,,  
 \end{align}
and
\begin{align}  \label{mesonIII}
  \langle \eta^{(\prime)}(p)   |   \bar q \, \gamma_5 \, q |  0 \rangle &\;= \;  - \frac{i}{2 \sqrt{2} m_q   }\, h_{\eta^{(\prime)} }^{q}    \,, \qquad   \, \, \,
  \langle \eta^{(\prime)} (p) |   \bar s \, \gamma_5 \, s |  0 \rangleÊ \;= \;  - \frac{i}{2 m_s   }\, h_{\eta^{(\prime)}}^{s}  \,,  \nonumber \\
  \langle \eta^{(\prime)}(p)   |   \bar q \, \gamma^{\mu} \gamma_5 \, q |  0 \rangle &\;= \; - \frac{i}{\sqrt{2}} f_{\eta^{(\prime)}}^{q} \, p^{\mu}  \,, \qquad   \, \, \,
  \langle \eta^{(\prime)} (p) |   \bar s \, \gamma^{\mu}  \gamma_5 \, s |  0 \rangleÊ \;= \;  - i f_{\eta^{(\prime)}}^{s} \, p^{\mu}    \,,   
 \end{align}
with $q=u,d$ and $ \hat m = (m_u + m_d)/2$ (we assume exact isospin symmetry).   Hadronic matrix elements for the gluonic operator $G_a^{\mu \nu}  \widetilde{G}_{\mu \nu}^{a} $ are similarly parametrized in terms of $a_{\eta^{(\prime)}}$:
\begin{align}  \label{omega}
 \langle  \eta^{(\prime)}(p)  |  \frac{ \alpha_s}{4 \pi}  \, G_a^{\mu \nu}  \widetilde{G}_{\mu \nu}^{a}   | 0  \rangle &=    a_{\eta^{(\prime)}}    \,.
  \end{align}
For the pion, $\langle  \pi(p)  |  \frac{ \alpha_s}{4 \pi}  \, G_a^{\mu \nu}  \widetilde{G}_{\mu \nu}^{a}   | 0  \rangle$ vanishes in the isospin limit ($m_u = m_d$) and is not considered here~\cite{Gross:1979ur}.   The axial anomaly of QCD relates the pseudoscalar and axial hadronic matrix elements with that of the gluonic operator,
\be \label{traceanomaly}
\partial_{\mu} (\bar q  \gamma^{\mu} \gamma_5 q) = 2 i m_q \bar q  \gamma_5 q + \frac{\alpha_s}{4 \pi } G_{\mu \nu}^{a} \widetilde{G}_{a}^{\mu \nu} \,,
\ee
implying the following relation among the parameters defined previously
\be \label{relations}
 a_{\eta^{(\prime)}}  = \frac{  f_{ \eta^{(\prime)} }^{q} m_{ \eta^{(\prime)} }^2 - h_{ \eta^{(\prime)} }^{q}  }{ \sqrt{2} } = f_{ \eta^{(\prime)} }^{s} m_{\eta^{(\prime)}}^2 - h_{ \eta^{(\prime)} }^{s} \,.
\ee
The pion decay constant is determined to be $f_{\pi} = 130.41  \pm 0.20$~MeV~\cite{Beringer:1900zz} while $a_{\eta} = 0.022 \pm 0.002$~GeV$^3$ and $a_{\eta^{\prime}} =0.056 \pm 0.002$~GeV$^3$~\cite{Feldmann:1998vh,Beneke:2002jn}.  Numerical values for the other parameters can be found in Table~\ref{tab:inputs}.  
For a recent analysis of the relevant  $\eta$ and $\eta'$ matrix elements within lattice QCD,  see Ref.~\cite{Michael:2013vba}.

 %%%%%%%%%%%%%%%%%%%%%%%%%%%%%%%%%%%%%%%%%%%%%%%%% Table: inputs  %%%%%%%%%%%%%%%%%%%%%%%%%%%%%%%%%%%%%%%%%%%%%%%%%%%
\begin{table}[ht!]
\setlength{\extrarowheight}{3pt}
\begin{center}
\caption{\it \small  Numerical values for the pseudoscalar and axial current densities, relevant for $\tau \rightarrow \mu P$ decays, obtained in the FKS mixing scheme~\cite{Feldmann:1998vh,Beneke:2002jn}. }
\vspace{0.4cm}
\doublerulesep 0.8pt \tabcolsep 0.2in
\begin{tabular}{|c|c||c|c| }
\hline 	
\rowcolor{RGray}
Pseudoscalar  & Value & Axial  & Value\\
\hline  \hline
$h_{\eta}^{q}$ &  $ 0.001 \pm 0.003$~GeV$^3$  & $f_{\eta}^{q}$ &  $ 0.11 \pm 0.01$~GeV   \\
\rowcolor{Gray}
$h_{\eta^{\prime}}^{q}$ & $ 0.001 \pm 0.002$~GeV$^3$  & $f_{\eta^{\prime}}^{q}$ &  $ 0.087 \pm 0.004$~GeV  \\
$h_{\eta}^{s}$ &  $ -0.055 \pm 0.003$~GeV$^3$ & $f_{\eta}^{s}$ &  $ -0.11 \pm 0.01$~GeV   \\
\rowcolor{Gray}
$h_{\eta^{\prime}}^{s}$ & $ 0.068 \pm 0.005$~GeV$^3$   & $f_{\eta^{\prime}}^{s}$ &  $ 0.135 \pm 0.006$~GeV  \\
\hline
\end{tabular}
\label{tab:inputs}
\end{center}\end{table}
%%%%%%%%%%%%%%%%%%%%%%%%%%%%%%%%%%%%%%%%%%%%%%%%%%%%%%%%%%%%%%%%%%%%%%%%%%%%%%%%%%%%%%%%%%%%%%%%%%%%%%%%%%%%%%%%%%%%%%%

%
%
%
%
\subsection{Semileptonic $\tau \rightarrow \mu \pi^+ \pi^-$ decays}
For the semileptonic decays $\tau \rightarrow \mu \pi^+ \pi^-$, the crucial point is that one needs a proper description of the hadronic dynamics for $\pi\pi$ invariant masses up to  $(m_{\tau}  - m_{\mu} )^2$.   Assuming isospin to be conserved, the hadronic matrix element for the vector current is given by
\be  \label{vectorff}
\langle   \pi^+(p_{\pi^+})   \pi^-(p_{\pi^{-}})   \bigl\vert  \frac{1}{2} \left(   \bar u \gamma^{\mu} u  - \bar d \gamma^{\mu} d   \right)     \bigr\vert   0 \rangle \;=  \; F_V(s) \left(    p_{\pi^+}  - p_{\pi^-} \right)^{\mu} \,,
\ee
where $F_V(s)$ denotes the pion vector form factor and $s = (p_{\pi^+} + p_{\pi^-})^2$ is the invariant mass of the pion pair.  This form factor can be determined phenomenologically by fitting the invariant mass distribution of $\tau \rightarrow \pi^- \pi^0 \nu_{\tau}$ decays using a dispersive parametrization, see Ref.~\cite{Celis:2013xja} and references therein.  

The hadronic matrix elements associated to scalar currents and the Parity-even gluonic operator $G_{\mu \nu}^{a}  G^{\mu \nu}_{a}$ are expressed in terms of the form factors $\Gamma_{\pi}(s), \Delta_{\pi}(s)$ and $\theta_{\pi}(s)$ respectively,
\begin{align}  \label{formfactors}
\langle \pi^{i}(p) \pi^{k}(p^{\prime}) | \theta_{\mu}^{\mu} |0 \rangle &= \theta_{\pi}(s) \delta^{ik} \,,  \\ \nonumber
\langle \pi^{i}(p) \pi^{k}(p^{\prime}) | m_u \bar u u + m_d \bar d d   | 0 \rangle &= \Gamma_{\pi}(s) \delta^{ik}   \,, \\ \nonumber 
\langle \pi^{i}(p) \pi^{k}(p^{\prime}) | m_s \bar s s  | 0 \rangle &= \Delta_{\pi}(s) \delta^{ik} \,.  
\end{align}
Here $\theta_{\mu}^{\mu}$ denotes the trace of the energy-momentum tensor given by
\be
\label{eq:QCDtrace}
\theta_{\mu}^{\mu} = - 9 \frac{\alpha_s}{8 \pi} \,G_{\mu \nu}^{a}  G^{\mu \nu}_{a}  + \sum_{q=u,d,s} m_q \,\bar q q ~,
\ee
where heavy quarks have been integrated out and the trace anomaly of the energy-momentum tensor has been taken into account.  The hadronic matrix element for the gluonic operator $G_{\mu \nu}^{a}  G^{\mu \nu}_{a}$ can then be written as
\be 
\langle \pi^{i}(p) \pi^{k}(p^{\prime}) |   \frac{ \beta_L}{4 \alpha_s} \,   G_{\mu \nu}^{a}  G^{\mu \nu}_{a}  |0 \rangle \; =  \, \left(   \theta_{\pi}(s) - \Gamma_{\pi}(s) - \Delta_{\pi}(s) \right)  \, \delta^{ik}   \; \,.
\ee
Here $\beta_L/(4 \alpha_s) = -9 \alpha_s/(8 \pi)$.   One can rely on a combination of dispersive methods and ChPT in order to obtain a reliable determination of the form factors $\{ \Gamma_{\pi}(s), \Delta_{\pi}(s), \theta_{\pi}(s)  \}$ in all the kinematical regime, these techniques were employed in Ref.~\cite{Donoghue:1990xh} to calculate the decay rate of a very light Higgs into two pions.  Recent works have used these methods for $\tau \rightarrow \ell \pi^+ \pi^-$ decays~\cite{Daub:2012mu,Celis:2013xja}, finding considerable improvements over previous treatments in the literature.    In this work we use the form factors determined in Ref.~\cite{Celis:2013xja}.

\end{appendix}

\subsection*{Acknowledgments}
%
%
%
%
%:
We would like to thank Kiyoshi Hayasaka for clarifications regarding the experimental limits on LFV $\tau$ decays and future prospects.   The work of A.C. has been supported in part by the Spanish Government and ERDF funds from the EU Commission [FPU grant AP2010-0308, Grants FPA2011-23778 and CSD2007-00042 (Consolider Project CPAN)] and by Generalitat Valenciana under Grant No. PROMETEOII/2013/007. 
The work of V.C. and E.P. is supported by the DOE   Office of Science, Nuclear Physics program. 

\bibliographystyle{h-physrev}
\bibliography{GLFV}

\begin{thebibliography}{10}

\bibitem{Raidal:2008jk}
M.~Raidal {\em et~al.},
\newblock Eur.Phys.J. {\bf C57}, 13 (2008), 0801.1826.

\bibitem{Cvetic:2002jy}
G.~Cvetic, C.~Dib, C.~Kim, and J.~Kim,
\newblock Phys.Rev. {\bf D66}, 034008 (2002), hep-ph/0202212.

\bibitem{Borzumati:1986qx}
F.~Borzumati and A.~Masiero,
\newblock Phys.Rev.Lett. {\bf 57}, 961 (1986).

\bibitem{Barbieri:1995tw}
R.~Barbieri, L.~J. Hall, and A.~Strumia,
\newblock Nucl.Phys. {\bf B445}, 219 (1995), hep-ph/9501334.

\bibitem{Hisano:1995cp}
J.~Hisano, T.~Moroi, K.~Tobe, and M.~Yamaguchi,
\newblock Phys.Rev. {\bf D53}, 2442 (1996), hep-ph/9510309.

\bibitem{Ellis:2002fe}
J.~R. Ellis, J.~Hisano, M.~Raidal, and Y.~Shimizu,
\newblock Phys.Rev. {\bf D66}, 115013 (2002), hep-ph/0206110.

\bibitem{Dedes:2002rh}
A.~Dedes, J.~R. Ellis, and M.~Raidal,
\newblock Phys.Lett. {\bf B549}, 159 (2002), hep-ph/0209207.

\bibitem{Masiero:2002jn}
A.~Masiero, S.~K. Vempati, and O.~Vives,
\newblock Nucl.Phys. {\bf B649}, 189 (2003), hep-ph/0209303.

\bibitem{Fukuyama:2003hn}
T.~Fukuyama, T.~Kikuchi, and N.~Okada,
\newblock Phys.Rev. {\bf D68}, 033012 (2003), hep-ph/0304190.

\bibitem{Brignole:2004ah}
A.~Brignole and A.~Rossi,
\newblock Nucl.Phys. {\bf B701}, 3 (2004), hep-ph/0404211.

\bibitem{Arganda:2008jj}
E.~Arganda, M.~Herrero, and J.~Portoles,
\newblock JHEP {\bf 0806}, 079 (2008), 0803.2039.

\bibitem{Herrero:2009tm}
M.~Herrero, J.~Portoles, and A.~Rodriguez-Sanchez,
\newblock Phys.Rev. {\bf D80}, 015023 (2009), 0903.5151.

\bibitem{Hirsch:2012ax}
M.~Hirsch, F.~Staub, and A.~Vicente,
\newblock Phys.Rev. {\bf D85}, 113013 (2012), 1202.1825.

\bibitem{Altmannshofer:2013lfa}
W.~Altmannshofer, R.~Harnik, and J.~Zupan,
\newblock JHEP {\bf 1311}, 202 (2013), 1308.3653.

\bibitem{Barenboim:1996vu}
G.~Barenboim and M.~Raidal,
\newblock Nucl.Phys. {\bf B484}, 63 (1997), hep-ph/9607281.

\bibitem{Cirigliano:2004mv}
V.~Cirigliano, A.~Kurylov, M.~Ramsey-Musolf, and P.~Vogel,
\newblock Phys.Rev. {\bf D70}, 075007 (2004), hep-ph/0404233.

\bibitem{Yue:2002ja}
C.-x. Yue, Y.-m. Zhang, and L.-j. Liu,
\newblock Phys.Lett. {\bf B547}, 252 (2002), hep-ph/0209291.

\bibitem{Diaz:2000cm}
R.~Diaz, R.~Martinez, and J.~A. Rodriguez,
\newblock Phys.Rev. {\bf D63}, 095007 (2001), hep-ph/0010149.

\bibitem{Kanemura:2005hr}
S.~Kanemura, T.~Ota, and K.~Tsumura,
\newblock Phys.Rev. {\bf D73}, 016006 (2006), hep-ph/0505191.

\bibitem{Paradisi:2005tk}
P.~Paradisi,
\newblock JHEP {\bf 0602}, 050 (2006), hep-ph/0508054.

\bibitem{Davidson:2010xv}
S.~Davidson and G.~J. Grenier,
\newblock Phys.Rev. {\bf D81}, 095016 (2010), 1001.0434.

\bibitem{Crivellin:2013wna}
A.~Crivellin, A.~Kokulu, and C.~Greub,
\newblock Phys.Rev. {\bf D87}, 094031 (2013), 1303.5877.

\bibitem{Celis:2013xja}
A.~Celis, V.~Cirigliano, and E.~Passemar,
\newblock Phys.Rev. {\bf D89}, 013008 (2014), 1309.3564.

\bibitem{Davidson:1993qk}
S.~Davidson, D.~C. Bailey, and B.~A. Campbell,
\newblock Z.Phys. {\bf C61}, 613 (1994), hep-ph/9309310.

\bibitem{Gabrielli:2000te}
E.~Gabrielli,
\newblock Phys.Rev. {\bf D62}, 055009 (2000), hep-ph/9911539.

\bibitem{Arnold:2013cva}
J.~M. Arnold, B.~Fornal, and M.~B. Wise,
\newblock Phys.Rev. {\bf D88}, 035009 (2013), 1304.6119.

\bibitem{McKeen:2012av}
D.~McKeen, M.~Pospelov, and A.~Ritz,
\newblock Phys.Rev. {\bf D86}, 113004 (2012), 1208.4597.

\bibitem{Redi:2013pga}
M.~Redi,
\newblock JHEP {\bf 1309}, 060 (2013), 1306.1525.

\bibitem{Ishiwata:2013gma}
K.~Ishiwata and M.~B. Wise,
\newblock Phys.Rev. {\bf D88}, 055009 (2013), 1307.1112.

\bibitem{Falkowski:2013jya}
A.~Falkowski, D.~M. Straub, and A.~Vicente,
\newblock (2013), 1312.5329.

\bibitem{delAguila:2008zu}
F.~del Aguila, J.~Illana, and M.~Jenkins,
\newblock JHEP {\bf 0901}, 080 (2009), 0811.2891.

\bibitem{Blanke:2009am}
M.~Blanke, A.~J. Buras, B.~Duling, S.~Recksiegel, and C.~Tarantino,
\newblock Acta Phys.Polon. {\bf B41}, 657 (2010), 0906.5454.

\bibitem{Aaij:2013fia}
LHCb collaboration, R.~Aaij {\em et~al.},
\newblock Phys.Lett. {\bf B724}, 36 (2013), 1304.4518.

\bibitem{Abe:2010sj}
Belle II Collaboration, T.~Abe,
\newblock (2010), 1011.0352.

\bibitem{Bobrov:2012kg}
A.~Bobrov and A.~Bondar,
\newblock Nucl.Phys.Proc.Suppl. {\bf 225-227}, 195 (2012), 1206.1909.

\bibitem{Raidal:1997hq}
M.~Raidal and A.~Santamaria,
\newblock Phys.Lett. {\bf B421}, 250 (1998), hep-ph/9710389.

\bibitem{Kuno:1999jp}
Y.~Kuno and Y.~Okada,
\newblock Rev.Mod.Phys. {\bf 73}, 151 (2001), hep-ph/9909265.

\bibitem{Kitano:2002mt}
R.~Kitano, M.~Koike, and Y.~Okada,
\newblock Phys.Rev. {\bf D66}, 096002 (2002), hep-ph/0203110.

\bibitem{Cirigliano:2009bz}
V.~Cirigliano, R.~Kitano, Y.~Okada, and P.~Tuzon,
\newblock Phys.Rev. {\bf D80}, 013002 (2009), 0904.0957.

\bibitem{deGouvea:2000cf}
A.~de~Gouvea, S.~Lola, and K.~Tobe,
\newblock Phys.Rev. {\bf D63}, 035004 (2001), hep-ph/0008085.

\bibitem{Black:2002wh}
D.~Black, T.~Han, H.-J. He, and M.~Sher,
\newblock Phys.Rev. {\bf D66}, 053002 (2002), hep-ph/0206056.

\bibitem{Dassinger:2007ru}
B.~Dassinger, T.~Feldmann, T.~Mannel, and S.~Turczyk,
\newblock JHEP {\bf 0710}, 039 (2007), 0707.0988.

\bibitem{Matsuzaki:2007hh}
A.~Matsuzaki and A.~Sanda,
\newblock Phys.Rev. {\bf D77}, 073003 (2008), 0711.0792.

\bibitem{Giffels:2008ar}
M.~Giffels, J.~Kallarackal, M.~Kramer, B.~O'Leary, and A.~Stahl,
\newblock Phys.Rev. {\bf D77}, 073010 (2008), 0802.0049.

\bibitem{Petrov:2013vka}
A.~A. Petrov and D.~V. Zhuridov,
\newblock Phys.Rev. {\bf D89}, 033005 (2014), 1308.6561.

\bibitem{Crivellin:2013hpa}
A.~Crivellin, S.~Najjari, and J.~Rosiek,
\newblock (2013), 1312.0634.

\bibitem{Daub:2012mu}
J.~Daub, H.~Dreiner, C.~Hanhart, B.~Kubis, and U.~Meissner,
\newblock JHEP {\bf 1301}, 179 (2013), 1212.4408.

\bibitem{Haber:2013mia}
H.~E. Haber,
\newblock (2013), 1401.0152.

\bibitem{Weinberg:1979sa}
S.~Weinberg,
\newblock Phys.Rev.Lett. {\bf 43}, 1566 (1979).

\bibitem{Buchmuller:1985jz}
W.~Buchmuller and D.~Wyler,
\newblock Nucl.Phys. {\bf B268}, 621 (1986).

\bibitem{Grzadkowski:2010es}
B.~Grzadkowski, M.~Iskrzynski, M.~Misiak, and J.~Rosiek,
\newblock JHEP {\bf 1010}, 085 (2010), 1008.4884.

\bibitem{Alonso:2013hga}
R.~Alonso, E.~E. Jenkins, A.~V. Manohar, and M.~Trott,
\newblock (2013), 1312.2014.

\bibitem{Alonso:2012pz}
R.~Alonso, M.~Gavela, L.~Merlo, S.~Rigolin, and J.~Yepes,
\newblock Phys.Rev. {\bf D87}, 055019 (2013), 1212.3307.

\bibitem{Pich:2012dv}
A.~Pich, I.~Rosell, and J.~J. Sanz-Cillero,
\newblock Phys.Rev.Lett. {\bf 110}, 181801 (2013), 1212.6769.

\bibitem{Contino:2013kra}
R.~Contino, M.~Ghezzi, C.~Grojean, M.~Muhlleitner, and M.~Spira,
\newblock JHEP {\bf 1307}, 035 (2013), 1303.3876.

\bibitem{Buchalla:2012qq}
G.~Buchalla and O.~Cata,
\newblock JHEP {\bf 1207}, 101 (2012), 1203.6510.

\bibitem{Buchalla:2013rka}
G.~Buchalla, O.~Cata, and C.~Krause,
\newblock Nucl.Phys. {\bf B880}, 552 (2014), 1307.5017.

\bibitem{Beringer:1900zz}
Particle Data Group, J.~Beringer {\em et~al.},
\newblock Phys.Rev. {\bf D86}, 010001 (2012).

\bibitem{Aubert:2009ag}
BaBar Collaboration, B.~Aubert {\em et~al.},
\newblock Phys.Rev.Lett. {\bf 104}, 021802 (2010), 0908.2381.

\bibitem{Hayasaka:2010np}
K.~Hayasaka {\em et~al.},
\newblock Phys.Lett. {\bf B687}, 139 (2010), 1001.3221.

\bibitem{Aubert:2006cz}
BaBar Collaboration, B.~Aubert {\em et~al.},
\newblock Phys.Rev.Lett. {\bf 98}, 061803 (2007), hep-ex/0610067.

\bibitem{Miyazaki:2007jp}
Belle Collaboration, Y.~Miyazaki {\em et~al.},
\newblock Phys.Lett. {\bf B648}, 341 (2007), hep-ex/0703009.

\bibitem{Miyazaki:2012mx}
Belle Collaboration, Y.~Miyazaki {\em et~al.},
\newblock Phys.Lett. {\bf B719}, 346 (2013), 1206.5595.

\bibitem{Miyazaki:2011xe}
Belle Collaboration, Y.~Miyazaki {\em et~al.},
\newblock Phys.Lett. {\bf B699}, 251 (2011), 1101.0755.

\bibitem{Miyazaki:2008mw}
Belle Collaboration, Y.~Miyazaki {\em et~al.},
\newblock Phys.Lett. {\bf B672}, 317 (2009), 0810.3519.

\bibitem{Sher:2003vi}
M.~Sher and I.~Turan,
\newblock Phys.Rev. {\bf D69}, 017302 (2004), hep-ph/0309183.

\bibitem{Kanemura:2004jt}
S.~Kanemura, Y.~Kuno, M.~Kuze, and T.~Ota,
\newblock Phys.Lett. {\bf B607}, 165 (2005), hep-ph/0410044.

\bibitem{Ahmed:1999gh}
CLEO Collaboration, S.~Ahmed {\em et~al.},
\newblock Phys.Rev. {\bf D61}, 071101 (2000), hep-ex/9910060.

\bibitem{Feldmann:1998vh}
T.~Feldmann, P.~Kroll, and B.~Stech,
\newblock Phys.Rev. {\bf D58}, 114006 (1998), hep-ph/9802409.

\bibitem{Beneke:2002jn}
M.~Beneke and M.~Neubert,
\newblock Nucl.Phys. {\bf B651}, 225 (2003), hep-ph/0210085.

\bibitem{Gross:1979ur}
D.~J. Gross, S.~Treiman, and F.~Wilczek,
\newblock Phys.Rev. {\bf D19}, 2188 (1979).

\bibitem{Michael:2013vba}
C.~Michael, K.~Ottnad, and C.~Urbach,
\newblock (2013), 1311.5490.

\bibitem{Donoghue:1990xh}
J.~F. Donoghue, J.~Gasser, and H.~Leutwyler,
\newblock Nucl.Phys. {\bf B343}, 341 (1990).

\end{thebibliography}

\end{document}